# Internal Heat and Energy Imbalance of Uranus

Xinyue Wang[1], Liming Li[2*], Michael Roman[3], Xi Zhang[4], Xun Jiang[1], Patrick M. Fry[5], Cheng Li[6], Gwenael Milcareck[7], Agustin Sanchez-Lavega[8], Santiago Perez-Hoyos[8], Ricardo Hueso[8], Tristan Guillot[9], Conor A. Nixon[10], Ulyana A. Dyudina[11], Robert A. West[12], Matthew E. Kenyon[12]

[1] Department of EAS, University of Houston, Houston, TX, USA, 77004.
[2] Department of Physics, University of Houston, Houston, TX, USA, 77004.
[3] University of Leicester, University Road, Leicester, UK, LE1 7RH.
[4] Department of Earth and Planetary Sciences, UCSC, Santa Cruz, CA, USA, 95064
[5] University of Wisconsin-Madison, Madison, WI, USA, 53706.
[6] University of Michigan, Ann Arbor, MI, USA, 48109.
[7] LMD/IPSL, Sorbonne Universit., PSL Research University, Paris, France, 75005.
[8] Departamento de Fisica Aplicada I, Escuela de Ingenieria UPV/EHU, Bilbao, Spain, 18013.
[9] Université Côte d'Azur, CNRS, Laboratoire Lagrange, Nice, France, 06108.
[10] NASA Goddard Space Flight Center, Greenbelt, MD, USA, 20771.
[11] Space Science Institute, Boulder, CO, USA, 80301.
[12] Jet Propulsion Laboratory, California Institute of Technology, Pasadena, CA, USA, 91109.

* To whom all correspondence should be addressed. E-mail: lli7@central.uh.edu

**With its extreme axial tilt, Uranus' radiant energy budget and internal heat remain among the most intriguing mysteries of our Solar System. Here, we present the global-average radiant energy budget spanning a complete orbital period (1946-2030), revealing significant seasonal variations driven primarily by Uranus' highly variable solar flux. Despite these fluctuations, emitted thermal power consistently exceeds absorbed solar power, indicating a net energy loss and ongoing global cooling. Based on the seasonal variations of radiant energy budget, we determine a statistically significant internal heat flux of $0.078\pm0.018$ Wm$^{-2}$, equivalent to $12.52\pm0.31\%$ of the absorbed solar power. This finding resolves a long-standing debate over whether Uranus possesses internal heat. We also examine the energy budget of Uranus' weather layer by combining the internal heat with the radiant energies, revealing significant energy imbalances at both global and hemispheric scales. The global average energy imbalance ranges from an excess of $9.5\pm3.9\%$ of the emitted power at perihelion to a deficit of $-8.3\pm3.7\%$ at aphelion. At the hemispheric scale, the imbalances are extreme, with the summer hemisphere exhibiting an energy excess of $95.0\pm5.4\%$ and the winter hemisphere experiencing a deficit of $-87.7\pm3.9\%$ relative to the emitted power. These global and hemispheric imbalances should be considered in theoretical and numerical models. The Uranus flagship mission, as recommended by the recent decadal survey, will provide crucial observations to address more unresolved questions and advance our understanding of this enigmatic ice giant.**

As an ice giant, Uranus has garnered special attention due to its unique obliquity and extremely strong seasonal variations. Compared to the gas giants Jupiter and Saturn, Uranus and its fellow ice giant, Neptune, have been visited by relatively few spacecraft, partly due to their long distance from Earth. This scarcity of observations is one of the main motivations behind the proposed flagship mission to Uranus, which has been ranked as the highest priority in NASA's planetary science program for the next decade, as reported in the Decadal Strategy for Planetary Science and Astrobiology 2023-2032 from the National Academies of Sciences, Engineering, and Medicine (National Academies Press, 2022).

The Voyager spacecraft, along with ground-based telescopes and Earth-orbiting observatories, has significantly enhanced our understanding of Uranus. However, many mysteries about this ice giant remain. In this study, we examine the radiant energy budget (REB) and internal heat of Uranus. The balance or imbalance between absorbed solar energy and emitted thermal energy, known as the REB, plays a crucial role in determining Uranus' thermal properties (e.g., Conrath et al., 1989; Hanel et al., 2003). This budget also regulates energy transfer and conversion within Uranus' atmospheric system, influencing its atmospheric dynamics and weather patterns. Giant planets generally possess internal heat, a quality closely tied to their formation and evolution (e.g., Smoluchowski, 1967; Hubbard, 1968; Salpeter, 1973; Flasar, 1973; Stevenson and Salpeter, 1977; Guillot, 2005). Although directly measuring internal heat is challenging, the REB provides an indirect yet essential method for estimating this property (e.g., Conrath et al., 1989).

A recent study (Wang et al., 2024) suggests that seasonal variations must be considered when investigating the REBs and internal heat of giant planets. Among all planets in our solar system, Uranus is expected to exhibit the strongest seasonal variations due to its unique obliquity (97.77°) and notable orbital eccentricity (0.047). These seasonal REB variations not only drive temporal changes in Uranus' atmospheric system but also play a critical role in determining its internal heat. However, studying Uranus' seasonal REB variations is challenging due to its long orbital period (~84 years) and extended seasonal durations (~20–22 years). Consequently, these variations remain largely unexplored, primarily due to the lack of long-term observations.

This limitation further complicates efforts to assess the planet's internal heat, leading to a discrepancy in its understanding. Observational analyses suggest that Uranus does not exhibit statistically significant internal heat (e.g., Pollack et al., 1986; Pearl et al., 1990), whereas detailed modeling and theoretical examination indicate that it should (e.g., Marley and McKay, 1999; Ge et al., 2023). Since other giant planets (i.e., Jupiter, Saturn, and Neptune) exhibit significant internal heat (Conrath et al., 1989; Ingersoll, 1990; Pearl and Conrath, 1991; Hanel et al., 2003; Li et al., 2018; Wang et al., 2024), resolving this discrepancy for Uranus is crucial. Determining whether Uranus has internal heat would provide key insights into the formation and evolution of giant planets.

Fortunately, many more observations have become available since the early studies, which were primarily based on Voyager data from the 1980s (Pollack et al., 1986; Pearl et al., 1990). In this study, we utilize Voyager data along with more recent observations to re-examine Uranus' REB and its seasonal variations, as well as to determine its internal heat. The detailed process of how the REB is measured based on available observations and studies are provided in the Supporting Information (SI) (Figs. S1-S18). Here, we briefly introduce the process. The basic methodology and observational data are summarized in Sections S1 and S2, respectively. This study emphasizes two key aspects: (1) seasonal variations and (2) analysis at both global and hemispheric scales. To examine a full seasonal cycle of Uranus' REB, we must consider an entire orbital period (~84 years). Modern observational records of Uranus date back to the 1950s, so we select the period from 1946 to 2030, which encompasses these observations from 1950 to the present. Uranus' REB is determined by two components: absorbed solar energy and emitted thermal energy. Absorbed solar power is generally calculated by subtracting reflected solar power from incident solar power at Uranus (i.e., the solar flux). We first reconstruct the solar power at Uranus at both global and hemispheric scales for the period 1946-2030 in Section S3 (Figs. S1-S4). With this known solar power, reflected solar power and hence absorbed solar power can be determined using the Bond albedo. Investigations of Uranus' Bond albedo and corresponding absorbed solar power at both global and hemispheric scales for 1946-2030 are presented in Section S4 (Figs. S5-S12). Uranus' emitted power at both global and hemispheric scales for the same period is discussed in Section S5 (Figs. S13-S17). Finally, we examine Uranus' REB, particularly in its two hemispheres, namely the Northern Hemisphere (NH) and the Southern Hemisphere (SH), in Section S6 (Fig. S18).

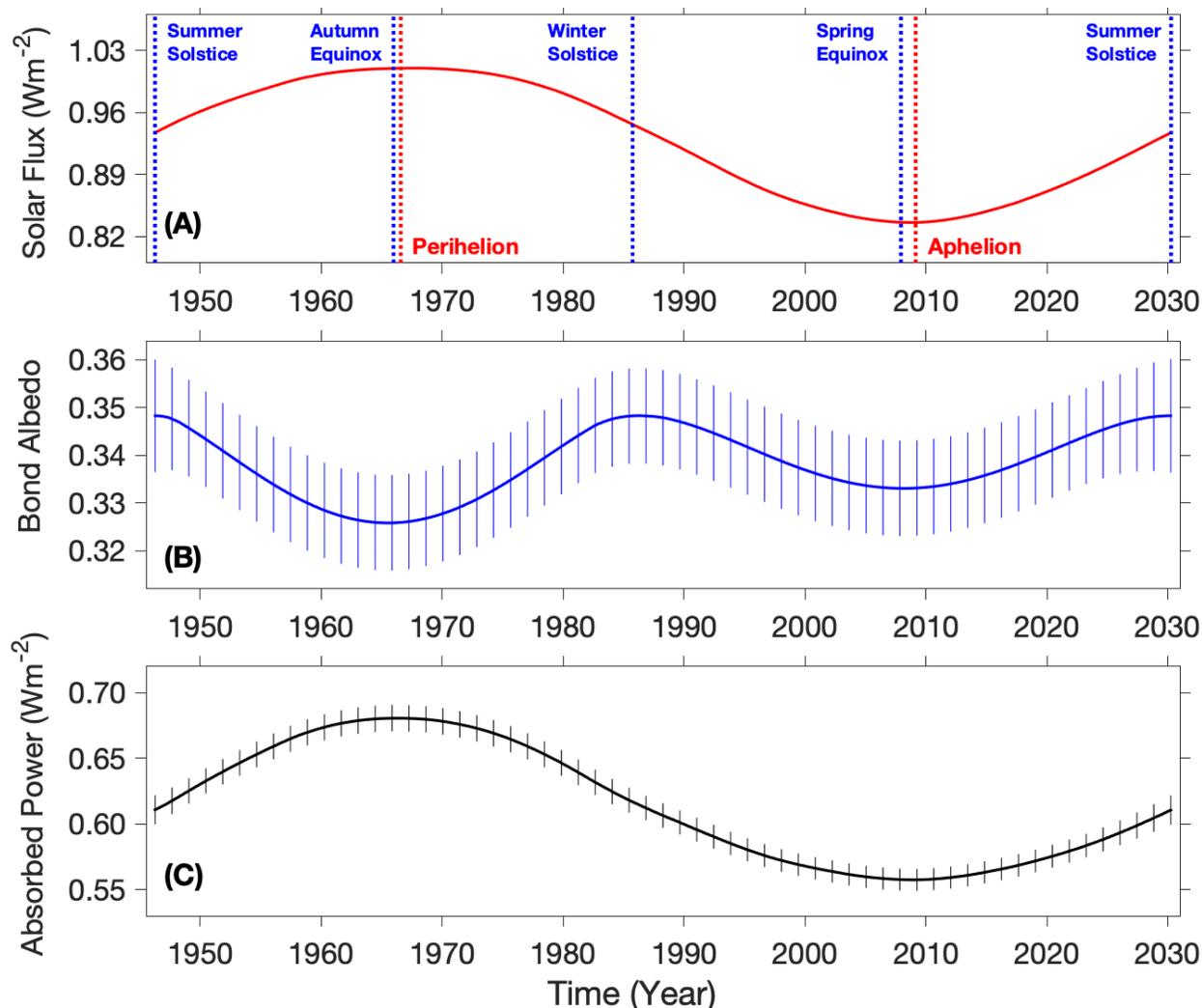

*Figure 1. Solar flux, Bond albedo, and absorbed solar power during Uranus' orbital period from 1946 to 2030.* (A) Global-average solar flux at Uranus. (B) Disk-average Bond albedo for the sunlit disk. (C) Global-average absorbed solar power. In panel A, the five blue vertical dashed lines mark the summer and winter solstices, as well as the spring and autumn equinoxes of the NH. The two red vertical dashed lines represent Uranus' perihelion and aphelion in its orbit around the Sun. The thin vertical solid lines in panels B and C indicate uncertainties in measurements.

Figure 1 presents the solar flux, Bond albedo, and absorbed solar power at the global scale. Uranus' relatively high orbital eccentricity causes the Sun-Uranus distance to increase by ~10.0% from perihelion in 1966 to aphelion in 2009, leading to a corresponding ~16.8% decrease in solar flux, from ~1.01 Wm$^{-2}$ at perihelion to ~0.84 Wm$^{-2}$ at aphelion (Figure 1A). Figure 1B indicates that the disk-average Bond albedo exhibits periodic variations, reaching maxima near the summer and winter solstices (1946, 1985, and 2030) and minima around the spring and autumn equinoxes (1966 and 2007) of each hemisphere. These variations likely result from both spatial differences in albedo (e.g., brighter polar regions compared to lower latitudes) (e.g., Lockwood, 1978; Karkoschka, 2001; Lockwood and Jerzykiewicz, 2006; Lockwood, 2019; Irwin et al., 2011, 2024) and seasonal changes in albedo (e.g., Rages et al., 2004; Hammel and Lockwood, 2007). The disk-average Bond albedo varies by ~6.3%, from a maximum of ~0.348 during the summer and winter solstices (1946, 1985, and 2030) to a minimum of ~0.326 during the autumn equinox (1966) of each hemisphere. Since seasonal variations in solar flux are stronger than those in Bond albedo, the seasonal pattern of the global-average absorbed solar power (Figure 1C) closely follows that of solar flux. Figure 1C further suggests that global-average absorbed power decreases by 18.2±2.0%, from 0.681±0.011 Wm$^{-2}$ around perihelion to 0.557±0.008 Wm$^{-2}$ around aphelion.

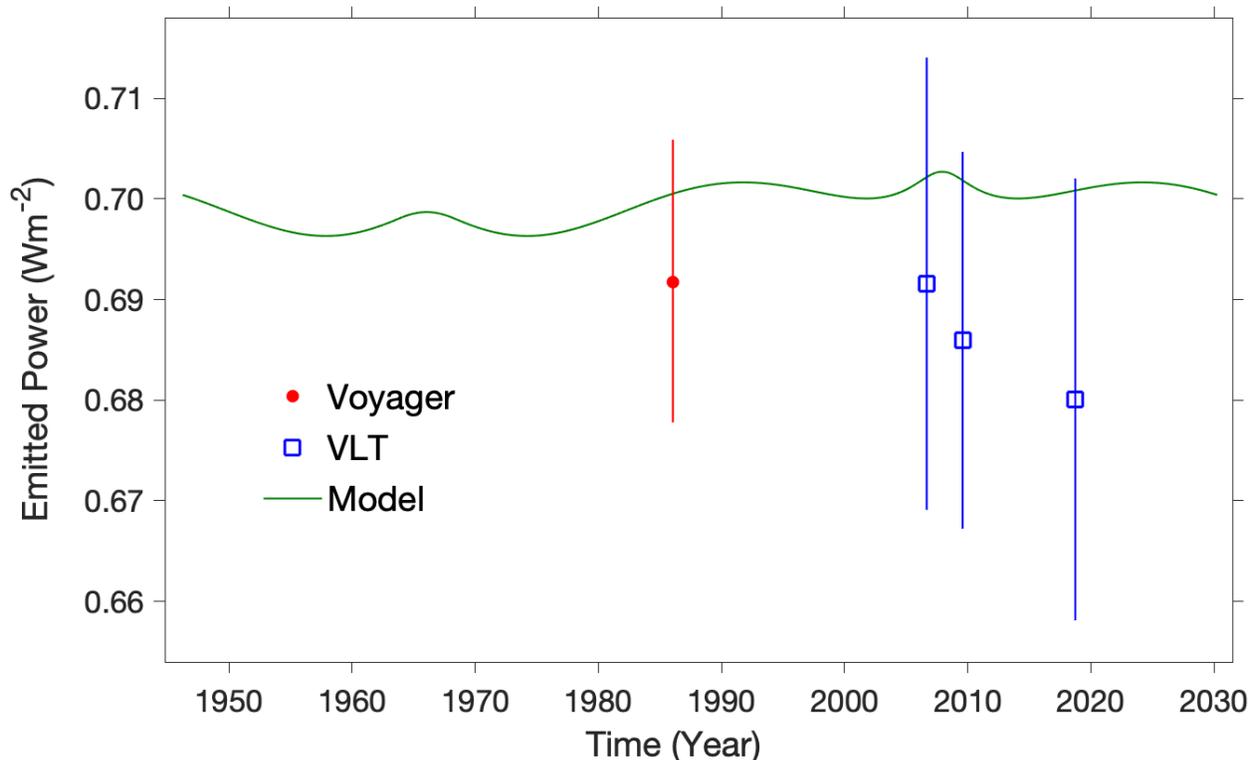

*Figure 2. Global-average emitted power of Uranus.* *The most accurate measurement based on Voyager observations in 1986 (Pearl et al., 1990) is shown as a red dot. New estimates of global-average emitted power for three years (2006, 2009, and 2018), derived from brightness temperature observations recorded by the Very Large Telescope (Orton et al., 2015; Roman et al., 2020), are represented by blue squares. The global-average emitted power obtained from numerical simulations of Uranus' atmosphere (Wallace, 1983) is shown as a green line. The vertical lines for the Voyager and VLT results indicate measurement uncertainties.*

Now, we discuss the other component of the REB: the emitted power. Figure 2 presents the best measurement of Uranus' global-average emitted power based on Voyager observations (Pearl et al., 1990), along with our estimates derived from Uranus' brightness temperature observations (see Section S5 in SI). Additionally, Figure 2 includes the global-average emitted power obtained from numerical simulations of Uranus' atmosphere (Wallace, 1983). First, the differences in emitted power between the Voyager measurement (0.692±0.014 Wm$^{-2}$ in 1986) and our estimates from brightness temperature (0.692±0.022 Wm$^{-2}$, 0.686±0.019 Wm$^{-2}$, and 0.680±0.022 Wm$^{-2}$ for 2006, 2009, and 2018 respectively) are smaller than their respective uncertainties, suggesting that temporal variations in Uranus' global-average emitted power are not statistically significant. The numerical simulations (Wallace, 1983) also show relatively small seasonal variations in the global-average emitted power, with a ratio of the standard deviation of seasonal variations to the annual mean of ~0.3%.

A comparison between the simulated emitted power and the best measurement from Voyager (Pearl et al., 1990) shows that the difference between them is smaller than the measurement uncertainty, indicating that the simulation and measurement are statistically consistent. More importantly, the seasonal variation in the simulated emitted power is smaller than the uncertainty of the Voyager measurement, further supporting the conclusion that Uranus' global-average emitted power does not exhibit significant seasonal variations. The radiative time constant of Uranus' atmosphere exceeds its orbital period (Conrath et al., 1990), which results in relatively weak seasonal variations in the large-scale thermal structure of Uranus' atmosphere (e.g., Wallace, 1983; Friedson and Ingersoll, 1987; Conrath et al., 1990; Bézard and Gautier, 1986; Bézard, 1990; Orton et al., 2015; Roman et al., 2020; Milcareck et al., 2024). Because the thermal structure is closely related to emitted power, these weak seasonal variations in thermal structure help explain the insignificant seasonal variations in Uranus' global-average emitted power.

In summary, both observations and simulations indicate that Uranus' global-average emitted power does not show statistically significant seasonal variations. Therefore, we assume it remains constant with season and adopt the best measurement from Voyager (0.692±0.014 Wm$^{-2}$) as the time-invariant value for our subsequent analysis of seasonal variations in the global-average REB.

To determine Uranus' REB and its seasonal variations, we compare absorbed solar power and emitted thermal power from 1946 to 2030, as shown in Figure 3. Panel A suggests that emitted thermal power exceeds absorbed solar power throughout Uranus' orbital period. Panel B shows that the difference between absorbed and emitted powers (i.e., absorbed power - emitted power) follows a similar pattern to absorbed solar power, since Uranus' global-average emitted power remains constant within measurement uncertainty (Figure 2). This difference ranges from 0.011±0.019 Wm$^{-2}$ at perihelion in 1966 to 0.134±0.018 Wm$^{-2}$ around aphelion in 2009. Correspondingly, the ratio of this power difference to emitted power increases from 1.6±2.8% in 1966 to 19.4±2.6% in 2009 (Figure 3, Panel C). Taking absorbed solar power as a reference, the ratio is even higher, rising from 1.63±2.87% at perihelion to 24.10±3.26% at aphelion.

For Uranus, the fact that emitted thermal power exceeds absorbed solar power indicates a radiant energy deficit, implying that Uranus is losing energy, though the cause and consequences of this energy loss are not yet fully understood. The difference between absorbed and emitted powers can be used to estimate internal heat (e.g., Conrath et al., 1989; Hanel et al., 2003). Previous estimates of Uranus' internal heat have not fully accounted for seasonal variations in the REB, leading to less precise estimates. This omission has contributed to discrepancies in past studies regarding whether Uranus has internal heat (Pollack et al., 1986; Pearl et al., 1990; Marley and McKay, 1999; Ge et al., 2023). Here, we re-examine Uranus' internal heat by accounting for seasonal variations in the REB.

The internal heat of giant planets is generally thought to be linked to planetary formation (e.g., Smoluchowski, 1967; Hubbard, 1968; Salpeter, 1973; Flasar, 1973; Stevenson and Salpeter, 1977; Guillot, 2005). Since the formation of giant planets occurs on timescales much longer than their orbital periods (e.g., Flasar, 1973), Uranus' internal heat should also have a timescale longer than its orbital period. In other words, we do not expect Uranus' internal heat to exhibit seasonal variations. Figure 3 shows that the difference between absorbed and emitted power varies across seasons, meaning that the power difference in any given season does not accurately represent Uranus' internal heat, which operates on a much longer timescale. To determine the internal heat, we average the power difference over a complete Uranus' year, yielding a value of 0.078±0.018 Wm$^{-2}$. The corresponding annual-mean absorbed and emitted powers are 0.614±0.012 Wm$^{-2}$ and 0.692±0.014 Wm$^{-2}$, respectively. Therefore, Uranus' internal heat accounts for 12.52±0.31% of absorbed solar power and 11.13±0.27% of emitted thermal power.

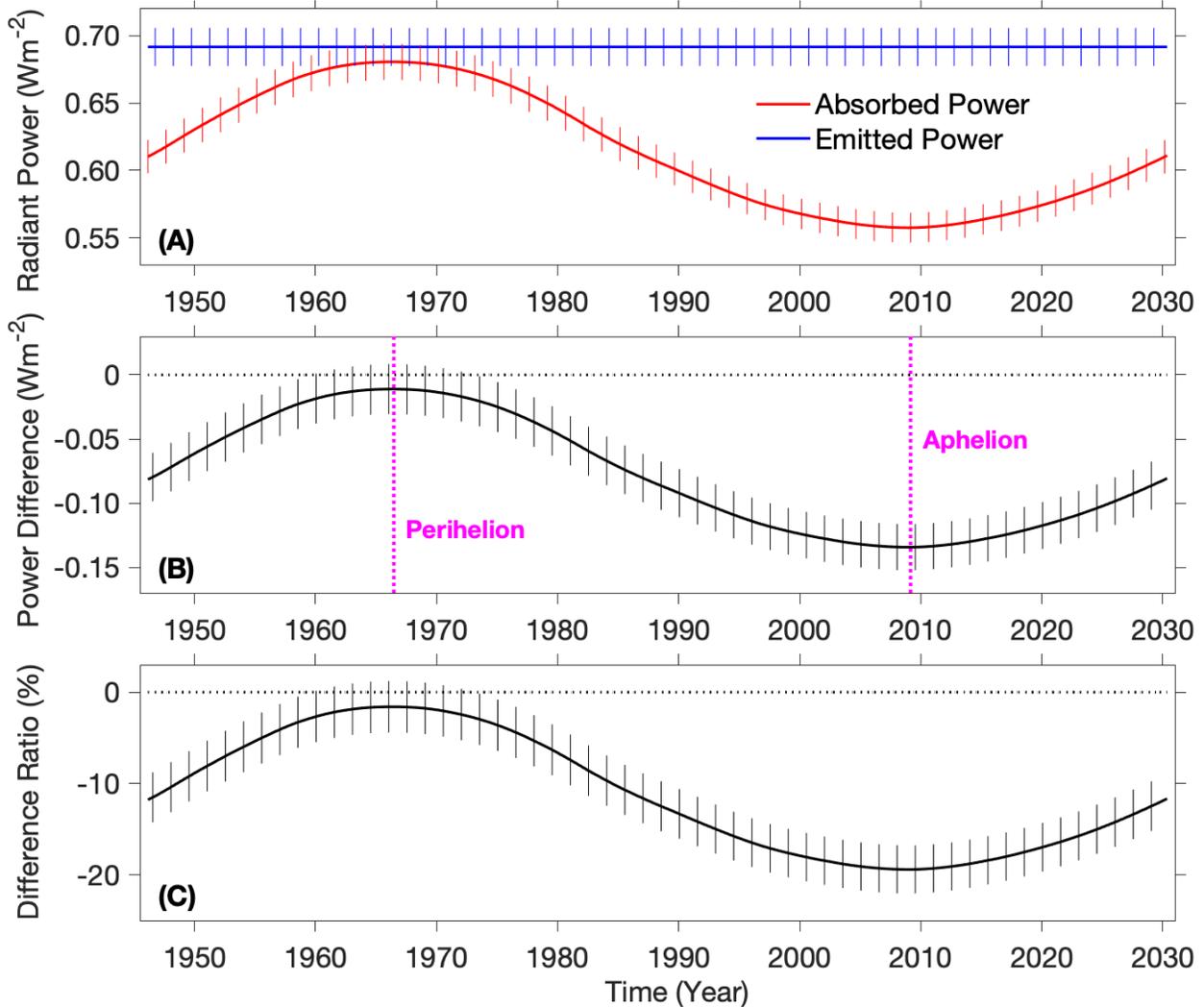

*Figure 3. Uranus' global-average absorbed power, emitted power, and their difference. (A) Comparison between global-average absorbed solar power and emitted thermal power. (B) Difference between absorbed and emitted powers (i.e., absorbed power minus emitted power). (C) Ratio of this difference to emitted power. The vertical thin solid lines in all three panels indicate uncertainties. The two magenta vertical dashed lines in panel B mark Uranus' perihelion and*

*aphelion in its orbit around the Sun. In panels B and C, the black horizontal dashed lines represent zero difference, serving as a reference.*

Thus, we conclude that Uranus does have internal heat, although the ratio of internal heat to radiant energy components is significantly lower than those of other giant planets (i.e., 113%, 139%, and 162% of absorbed solar power for Jupiter, Saturn, and Neptune, respectively) (Pearl and Conrath, 1991; Li et al., 2018; Wang et al., 2024). Various theories have been proposed to explain Uranus' relatively low internal heat, including hypotheses regarding its interior structure and evolutionary history (Hubbard, 1978, 1980; Stevenson, 1987; Podolak et al., 1991). The seasonal variations in the REB and new measurements of internal heat presented here provide additional constraints and insights to refine models of Uranus' interior and evolutionary processes.

The internal heat also plays an important role in the atmospheric systems of giant planets. It is one of the driving forces of the weather layer of giant planets, which is defined as the upper troposphere including clouds. For Uranus' weather layer, which ranges from the tropopause at ~ 100 mbar (West et al., 1991) to the lower boundary near the water clouds at ~ 100 bars (Allison et al., 1991), we can combine the seasonally varying absorbed solar power and the internal heat (0.078±0.018 Wm$^{-2}$) as the input power. In contrast, the output power is determined solely by the emitted thermal power.

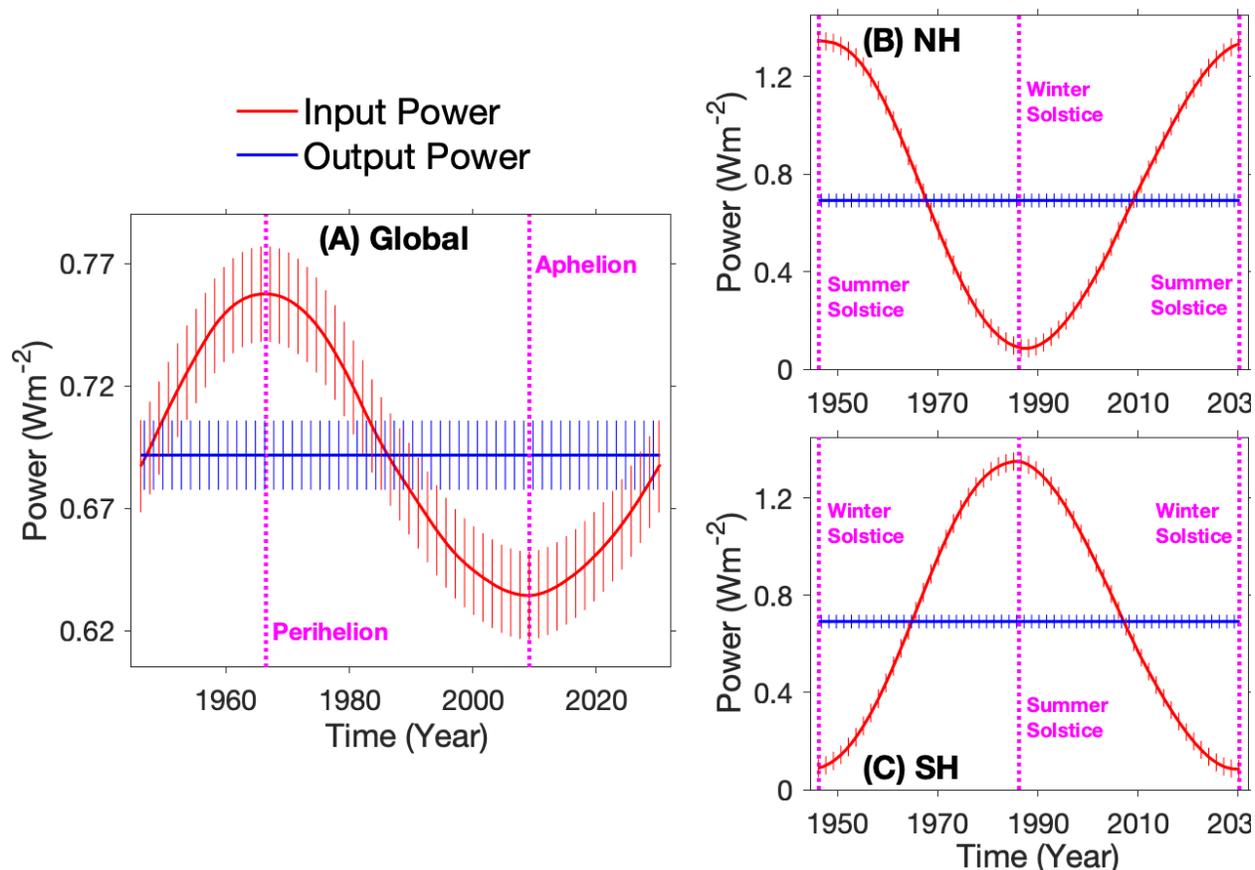

*Figure 4. Energy budget of Uranus' weather layer at global and hemispheric scales. (A) Comparison of global-average input and output powers. (B) and (C) show the same comparison as (A) but for the NH and SH, respectively. The input power is determined by combining the absorbed solar power with the internal heat, as described in the text. The output power is derived*

*solely from the emitted thermal power. In panel A, the two magenta vertical dashed lines indicate Uranus' perihelion and aphelion in its orbit around the Sun. In panels B and C, the magenta vertical dashed lines mark the summer and winter solstices for the respective hemisphere. The thin vertical solid lines in panels B and C represent measurement uncertainties.*

Panel A of Fig. 4 compares the input and output powers on a global scale, revealing a significant energy imbalance. The global-average energy imbalance ranges from an excess of 9.5±3.5% of the emitted power at perihelion in 1966 to a deficit of -8.3±3.3% of the emitted power at aphelion in 2009. Figure 4 also shows that the global-average energy excess primarily occurs in the two seasons around perihelion (i.e., summer and autumn in the NH), while the global-average energy deficit mainly appears in the two seasons around aphelion (i.e., winter and spring in the NH). The strong seasonal variations in absorbed flux, which are much greater than the nearly constant emitted power, result in a global-average seasonal energy imbalance.

Uranus' extreme axial tilt (97.7°) is expected to cause pronounced seasonal variations in the hemispheric-average energy budget. Therefore, we also examine the energy budget of the weather layer at the hemispheric scale. Assuming that the internal heat does not differ between hemispheres, we use the global-average internal heat to approximate the hemispheric-average internal heat. Then we can combine the hemispheric-average internal heat with the hemispheric REB (Fig. S18) to examine the energy budget of the weather layer at the hemispheric scale. Panels B and C of Fig. 4 illustrate the hemispheric energy budget of Uranus' weather layer, which exhibits extreme energy imbalances on a seasonal scale. The two hemispheres experience completely opposite energy imbalances at the solstices: one hemisphere undergoes an extreme energy excess, while the other experiences a significant energy deficit. The ratio of the hemispheric-average power difference (i.e., input power - output power) to output power varies substantially, from 95.0±6.7% at the summer solstice to -87.7±3.9% at the winter solstice in each hemisphere. If we take the input power as reference, the energy deficit at the winter solstice of each hemisphere can reach ~ -713.1±164.3% of the input power.

Although the two hemispheres experience entirely opposite energy imbalances at the solstices (Fig. 4), the thermal structure of Uranus' upper troposphere and lower stratosphere does not exhibit a pronounced asymmetry between them (e.g., Conrath et al., 1990, 1991; Orton et al., 2015; Roman et al., 2020). Therefore, a mechanism is required to transport solar heating from the sunlit hemisphere to the other hemisphere during solstices, possibly related to meridional circulation in the upper atmosphere (e.g., Friedson and Ingersoll, 1987). The quantitative characteristics of energy imbalance at both global and hemispheric scales can help constrain the mechanisms responsible for solar heating transport on Uranus.

In this study, Uranus' Bond albedo is determined by measuring the geometric albedo across a complete wavelength range and re-examining the phase function. The Bond albedo measurements are then used to determine seasonal variations in absorbed solar power. Seasonal variations in Uranus' emitted power are also examined through both observations and simulations. Based on these measurements, we provide the first global-average picture of Uranus' REB over a complete orbital period (1946-2030). While the thermal structure and emitted power exhibit little seasonal variation due to the long radiative timescale, the global-average REB undergoes significant seasonal changes, primarily driven by Uranus' strongly variable seasonal solar flux. Despite these fluctuations, the emitted thermal power consistently exceeds the absorbed solar power, indicating that Uranus is losing energy and undergoing global cooling.

Seasonal variations in the REB, which were not fully accounted for in previous estimates of internal heat, allow for a refinement of the internal heat. Our new analysis reveals a statistically significant internal heat of 0.078±0.018 Wm$^{-2}$, which constitutes 12.52±0.31% of the absorbed solar power. This determination resolves a long-standing debate about whether Uranus possesses internal heat. However, the ratio of internal heat to absorbed solar power is significantly lower than these of other giant planets. The quantitative characteristics of Uranus' internal heat provide observational constraints that can be used to develop the theories of planetary formation for giant planets, including Uranus. For Uranus' atmosphere, previous numerical models either omitted internal heat or incorporated it with imprecise values (e.g., Wallace, 1983; Bézard and Gautier, 1985; Friedson and Ingersoll, 1987; Bézard, 1990; Conrath et al., 1990; Milcarek et al., 2024). The new measurements of internal heat offer a basis for re-evaluating these atmospheric models.

Some theoretical studies (Wallace, 1983; Bézard and Gautier, 1985; Friedson and Ingersoll, 1987; Bézard, 1990) suggest that internal heat may vary with latitude and season if it is associated more with the weather layer rather than originating from the planet's deep interior. If we define the difference between absorbed and emitted powers at any given time as instantaneous internal heat, we find that the global-scale instantaneous internal heat varies from 1.60±2.82% of the emitted power at perihelion to 19.42±2.63% at aphelion. The hemispheric-scale REB suggests that instantaneous internal heat can be much larger at the hemispheric scale than at the global scale. In fact, the hemispheric-average instantaneous internal heat can reach ~100% of the emitted power during solstice seasons, comparable to the global-scale internal heat of other giant planets (Jupiter, Saturn, and Neptune).

Taking internal heat into account provides a more complete picture of the energy budget for Uranus' weather layer. The energy budget of Uranus exhibits significant imbalances at both global and hemispheric scales. In particular, the hemispheric-average energy budget shows an extreme energy excess of 95.0±5.4% of the emitted power at the summer solstice and a strong energy deficit of -87.7±3.9% at the winter solstice in each hemisphere. These extreme hemispheric-scale energy imbalances are primarily due to Uranus' unique axial tilt (97.7°). While numerical models (e.g., Wallace, 1983; Bézard and Gautier, 1985; Friedson and Ingersoll, 1987; Bézard, 1990; Conrath et al., 1990; Milcarek et al., 2024) have greatly advanced our understanding of Uranus' atmospheric system, some key observational characteristics (e.g., tropical warming in the upper troposphere) remain poorly simulated. A more complete picture of the energy budget may provide new insights for future numerical modeling efforts.

While current investigations help us better understand Uranus' REB and internal heat, more fundamental questions remain unanswered. For example, the meridional distribution of the REB, which plays a critical role in the large-scale circulation of planetary atmospheres, has not yet been determined for Uranus. Ground-based telescopes and Earth-orbiting observatories can observe Uranus' Earth-facing hemisphere, but these observations are generally limited in viewing geometry and latitude coverage. More importantly, the great distance between Earth and Uranus results in poor spatial resolution, making it difficult to resolve the planet's meridional energy distribution. On the other hand, the only flyby observations, conducted by Voyager, also had limitations in viewing geometry and wavelength coverage, restricting their ability to examine Uranus' meridional energy distribution. The Uranus orbiter mission recommended by the recent decadal survey presents a valuable opportunity to investigate the planet's REB and its spatiotemporal variability. By carefully selecting observation wavelengths and planning the spacecraft's orbit, this flagship mission will not only help answer fundamental questions about Uranus' REB but also significantly advance our understanding of other aspects of this ice giant.

**References**

Allison, M., Beebe, R. F., Conrath, B. J., Hinson, D. P. and Ingersoll, A. P., 1991. Uranus atmospheric dynamics and circulation. In Uranus. The University of Arizona Press.

Bézard, B., 1990. Seasonal thermal structure of the atmospheres of the giant planets. Advances in space research 10, 89-98.

Bezard, B. and Gautier, D., 1985. A model of the spatial and temporal variation of the Uranus thermal structure. In The Jovian Atmospheres. NASA Conference Publication 2441.

Conrath, B. J., Hanel, R. A. & Samuelson, R. E, 1989. Thermal Structure and Heat Balance of the Outer Planets. In Origin and Evolution of Planetary and Satellite Atmospheres. The University of Arizona Press.

Conrath, B.J., Gierasch, P.J. and Leroy, S.S., 1990. Temperature and circulation in the stratosphere of the outer planets. Icarus 83, 255-281.

Flasar, F. M., 1973. Gravitational energy sources in Jupiter. Astrophys. J. 186, 1097-1106.

Friedson, J. and Ingersoll, A.P., 1987. Seasonal meridional energy balance and thermal structure of the atmosphere of Uranus: A radiative-convective-dynamical model. Icarus 69, 135-156.

Ge, H., Li, C., Zhang, X. and Moeckel, C., 2024. Heat-flux-limited Cloud Activity and Vertical Mixing in Giant Planet Atmospheres with an Application to Uranus and Neptune. The Planetary Science Journal 5, 101.

Guillot, T. 2005. The interiors of giant planets: Models and outstanding questions. Annu. Rev. Earth Planet. Sci. 33, 493-530.

Hammel, H.B. and Lockwood, G.W., 2007. Long-term atmospheric variability on Uranus and Neptune. Icarus 186, 291-301.

Hanel, R. A., Conrath, B. J., Jennings, D. E. & Samuelson, R. E., 2003. Exploration of the Solar System by Infrared Remote Sensing. Cambridge Univ. Press.

Hubbard, W. B., 1968. Thermal structure of Jupiter. The Astrophysical Journal 152, 745-754.

Ingersoll, A. P., 1990. Atmospheric dynamics of the outer planets. Science 248, 308-315.

Irwin, P. G. J., Teanby, N. A., Davis, G. R., Fletcher, L. N., Orton, G. S., Tice, D. and Kyffin, A., 2011. Uranus' cloud structure and seasonal variability from Gemini-North and UKIRT observations. Icarus 212, 339-350.

Irwin, P.G., Dobinson, J., James, A., Teanby, N.A., Simon, A.A., Fletcher, L.N., Roman, M.T., Orton, G.S., Wong, M.H., Toledo, D. and Pérez-Hoyos, S., 2024. Modelling the seasonal cycle of Uranus' colour and magnitude, and comparison with Neptune. Monthly Notices of the Royal Astronomical Society 527, 11521-11538.

Karkoschka, E., 2001. Uranus' apparent seasonal variability in 25 HST filters. Icarus 151, 84-92.

Li, L., X. Jiang, R. A. West, P. J. Gierasch, S. Perez-Hoyos, A. Sanchez-Lavega, L. N. Fletcher, J. J. Fortney, B. Knowles, C. C. Porco, K. H. Baines, P. M. Fry, A. Mallama, R. K. Achterberg, A. A. Simon, C. A. Nixon, G. S. Orton, U. A. Dyudina, S. P. Ewald, 2018. Less absorbed solar energy and more internal heat for Jupiter. Nature Communications, doi:10.1038/s41467-018-06107-2.

Lockwood, G.W., 1978. Analysis of photometric variations of Uranus and Neptune since 1953. Icarus 35, 79-92.

Lockwood, G. W. and Jerzykiewicz, M., 2006. Photometric variability of Uranus and Neptune, 1950–2004. Icarus 180, 442-452.

Lockwood, G. W., 2019. Final compilation of photometry of Uranus and Neptune, 1972–2016. Icarus 324, 77-85.

Marley, M. S. & McKay, C. P., 1999. Thermal structure of Uranus' atmosphere. Icarus 138, 268-286.

Milcareck, G., Guerlet, S., Montmessin, F., Spiga, A., Leconte, J., Millour, E., Clement, N., Fletcher, L.N., Roman, M.T., Lellouch, E., Moreno, R., Cavalie, T., Carrion-Gonzalez, O., 2024. Radiative-convective models of the atmospheres of Uranus and Neptune: heating sources and seasonal effects. Astronomy & Astrophysics 686, A303.

National Academies of Sciences, Engineering, and Medicine, 2022. Origins, Worlds, and Life: A Decadal Strategy for Planetary Science and Astrobiology 2023-2032. The National Academies Press.

Orton, G.S., Fletcher, L.N., Encrenaz, T., Leyrat, C., Roe, H.G., Fujiyoshi, T. and Pantin, E., 2015. Thermal imaging of Uranus: upper-tropospheric temperatures one season after Voyager. Icarus 260, 94-102.

Pearl, J. C., Conrath, B. J., Hanel, R. A., Pirraglia, J. A., Coustenis, A., 1990. The albedo, effective temperature, and energy balance of Uranus, as determined from Voyager IRIS data. Icarus 84, 12-28.

Pearl, J. C. & Conrath, B. J., 1991. The albedo, effective temperature, and energy balance of Neptune, as determined from Voyager data. Journal of Geophysical Research 96, 18921-18930.

Pollack, J. B., Rages, K., Baines, K. H., Bergstralh, J. T., Wenkert, D. & Danielson, G. E. 1986. Estimates of the bolometric albedos and radiation balance of Uranus and Neptune. Icarus 65, 442-466.

Rages, K. A., Hammel, H. B. and Friedson, A. J., 2004. Evidence for temporal change at Uranus' south pole. Icarus 172, 548-554.

Roman, M.T., Fletcher, L.N., Orton, G.S., Rowe-Gurney, N. and Irwin, P.G., 2020. Uranus in northern midspring: persistent atmospheric temperatures and circulations inferred from thermal imaging. The Astronomical Journal 159, 45.

Salpeter, E., 1973. On convection and gravitational layering in Jupiter and stars of low mass. Astrophys 181, L83-L86.

Smoluchowski, R., 1967. Internal structure and energy emission of Jupiter. Nature 215, 691-695.

Stevenson, D. J. & Salpeter, E. E., 1977. The dynamics and helium distributions in hydrogen-helium planets. Astrophys. J. Suppl. 35, 239-261.

Wang, X., L. Li, X. Jiang, P. M. Fry, R. A. West, C. A. Nixon, L. Guan, T. D. Karandana, R. Albright, J. E. Colwell, T. Guillot, M. D. Hofstadter, M. E. Kenyon, A. Mallama, S. Perez-Hoyos, A. Sanchez-Lavega, A. Simon, D. Wenkert, X. Zhang, Cassini spacecraft reveals global energy imbalance of Saturn. Nature Communications, doi:10.1038/s41467-024-48969-9, 2024.

Wallace, L., 1983. The seasonal variation of the thermal structure of the atmosphere of Uranus. Icarus 54, 110-132.

West, R.A., Baines, K.H. and Pollack, J.B., 1991. Clouds and aerosols in the Uranian atmosphere. In Uranus. The University of Arizona Press.

# Supporting Information for
# Internal Heat and Energy Imbalance of Uranus

**S1. Methodology**

The REB (REB) of Uranus is determined by two components: the emitted thermal power and the absorbed solar power. The methodology for computing the two components for planets has been well described in previous literature (Conrath et al., 1989; Hanel et al., 2003) and in our investigations of the REBs of Jupiter and Saturn (Li et al., 2010, 2012, 2018; Wang et al., 2024). Here, we briefly introduce the methodology.

To determine Uranus' absorbed solar energy, we calculate it by subtracting the reflected solar power from the incident solar power. The reflected solar radiance is typically represented by the Bond albedo, which is defined as the ratio of reflected solar power (i.e., the integral of reflected solar radiance over wavelength and phase angle) to the incident solar power (e.g., Li et al., 2023). The phase angle is the angle between the line connecting Uranus to the Sun and the line connecting Uranus to an observer. Thus, the key parameter for measuring absorbed solar power is the Bond albedo, which can be expressed as the product of the geometric albedo and the phase integral. The geometric albedo represents the reflectivity at 0° phase angle, where reflectivity is defined as the ratio of reflected solar radiance to incident solar radiance at any given phase angle (Li et al., 2023). The ratio between the reflected solar radiance at different phase angles and the reflected radiance at 0° phase angle is referred to as the phase function. The integral of the phase function over the phase angle is the phase integral (Conrath et al., 1989; Li et al., 2023).

The fundamental approach of computing a planet's emitted power is to integrate the outgoing thermal radiance over both wavelength and emission angle. The emission angle is defined as the angle between the surface normal vector at a point on a planet and the vector to an observer. For Uranus, the emission angle coverage is highly limited for most available observations. As a result, direct measurements of Uranus' emitted power through integration of outgoing thermal radiance over the emission angle are not feasible in most cases. In this study, we develop a new method to estimate Uranus' emitted power, as detailed in Section S5. Once the two radiant energy components are determined, Uranus' REB can be evaluated by comparing them. Furthermore, the difference between these two energy components provides an estimate of Uranus' internal heat (Conrath et al., 1989; Hanel et al., 2003).

**S2. Summary of Observational Datasets and Model Simulations**

As discussed in the previous section, determining the absorbed solar power of Uranus requires knowledge of its incident solar power and Bond albedo. To determine the emitted thermal power, observations of outgoing radiance are necessary. Ideally, the optical and thermal observations used to measure Uranus' absorbed and emitted powers should have sufficient wavelength coverage (i.e., visible wavelengths for absorbed solar power and infrared wavelengths for emitted thermal power) and viewing geometry (e.g., phase angle for reflected solar power and emission angle for emitted thermal power). Therefore, we search for high-quality observational data at both visible and infrared wavelengths, along with corresponding published studies, to examine Uranus' REB and internal heat as comprehensively as possible. Compared to optical observations of Uranus in visible wavelengths, far-infrared observations are relatively scarce. In particular, observations of the radiance at wavelengths around 50 μm (i.e., the radiance that dominates emitted thermal power) are very limited. As a result, numerical model simulations of outgoing thermal emission are also included in our analysis of Uranus' emitted thermal power.

Uranus exhibits extreme seasonal variations in the spatial distribution of solar flux due to its unique axial tilt (97.77°) (see the next section). A recent study (Wang et al., 2024) suggests that seasonal variations of solar flux play a critical role in the REB and internal heat of giant planets. Therefore, investigating the seasonal variations in Uranus' REB is a primary focus of this study. Given Uranus' long orbital period (~84 years) and the availability of high-quality observations primarily beginning in the 1950s, we select the orbital period from 1946 to 2030 for analysis. This period includes the observational record from 1950 to the present and represents a complete Uranian year, as both 1946 and 2030 correspond to the summer solstice of the NH.

## S3. Incident Solar Power During the Period of 1946-2030

The incident solar power per unit area at Uranus, also referred to as solar flux, serves as the basis for computing Uranus' Bond albedo and its absorbed solar power. We first use the datasets collected on Earth (available from the Solar Radiation and Climate Experiment (SORCE) at http://lasp.colorado.edu/lisird/data/sorce_ssi_l3/) to obtain the solar flux for Earth from 1946 to 2019. Given that solar flux varies by less than 1% over time, we assume it remains constant from 2019 to 2030 (see Panel A of Fig. S1). Using the solar flux at Earth and the Sun-Uranus distance, we compute the solar flux at Uranus. The Sun-Uranus distance for 1946-2030 is obtained from NASA/JPL's Solar System Dynamics Horizons Web Interface (https://ssd.jpl.nasa.gov/horizons.cgi), as shown in Panel B of Fig. S1. Uranus' relatively large orbital eccentricity (0.0469) causes the Sun-Uranus distance to increase by ~10.0%, from ~18.28 AU at perihelion in 1966 to ~20.10 AU at aphelion in 2009. Correspondingly, the solar flux at Uranus decreases by ~17.2%, from ~4.07 Wm$^{-2}$ at perihelion to ~3.37 Wm$^{-2}$ at aphelion (Panel C of Fig. S1).

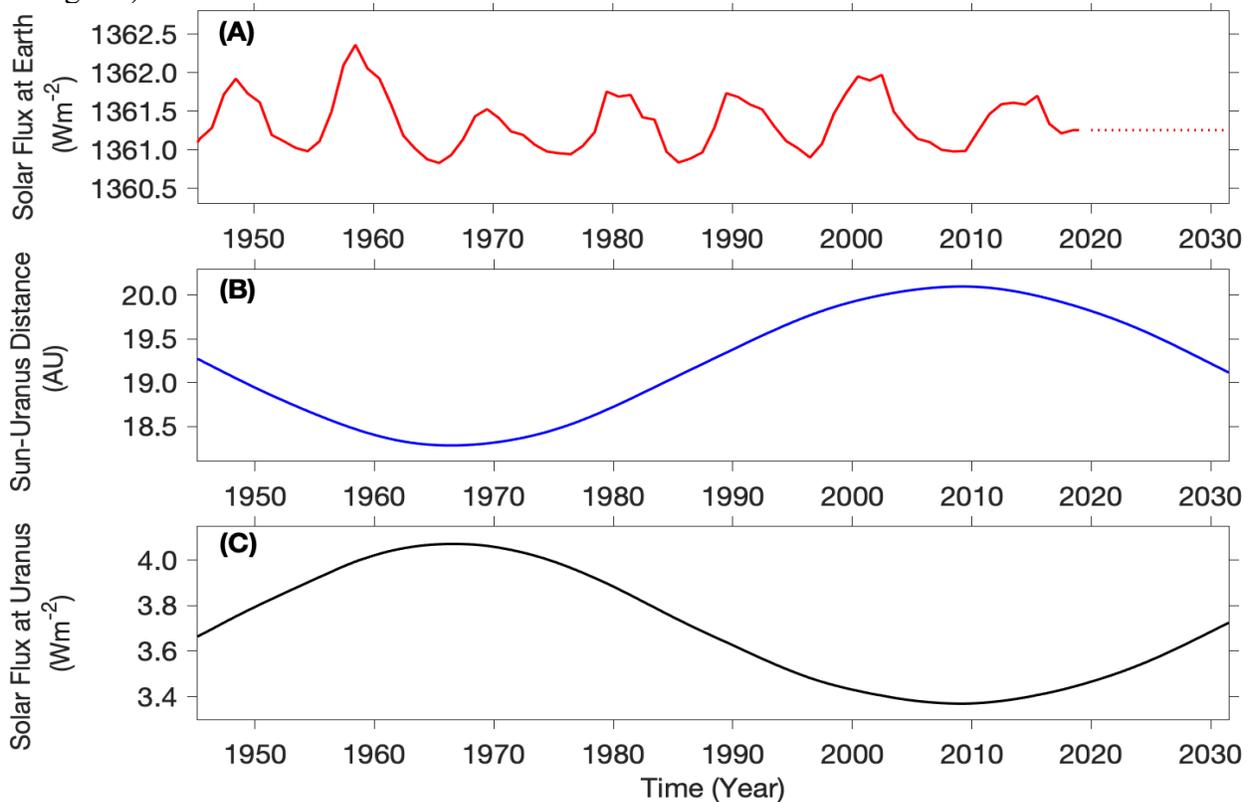

***Figure S1. Solar flux at Earth and Uranus from 1946 to 2030.*** *(A) Solar flux at Earth's distance. (B) Distance between the Sun and Uranus. (C) Solar flux at Uranus' distance. The solar flux at*

*Earth from 1946 to 2019 is based on observational data from SORCE, while the values for 2019-2030 are assumed to remain the same as in 2019. The Sun-Uranus distance is obtained from the JPL Horizons system.*

The solar flux at Uranus can be further used to calculate Uranus' global-average solar flux, in which the ratio of the global surface area to the solar-illuminated disk area is considered. For a sphere, the ratio between the global area ($4\pi r^2$, where r is the radius of the sphere) and the disk area ($\pi r^2$) is 4. Uranus is an oblate planet, and the ratio of global area to solar-illuminated disk area varies around 4. The solar-illuminated disk area can be approximated by an ellipse with a short semiaxis $r_s$ and a long semiaxis $r_l$. The long semiaxis $r_l$ is the equatorial radius of Uranus ($r_{eq}$ =25559 km) and the short semiaxis $r_s$ can be expressed as $(r_{eq} \times r_{po})/\sqrt{\left[(r_{eq}\cos{(lat_{ss})})^2 + (r_{po}\sin{(lat_{ss})})^2\right]}$, where $r_{po}$ is the polar radius of Uranus and $lat_{ss}$ is the sub-solar latitude. Then we can calculate the area of the ellipse for the solar-illuminated disk area. The global surface area of Uranus can be computed by following the equation of the surface area of an oblate spheroid.

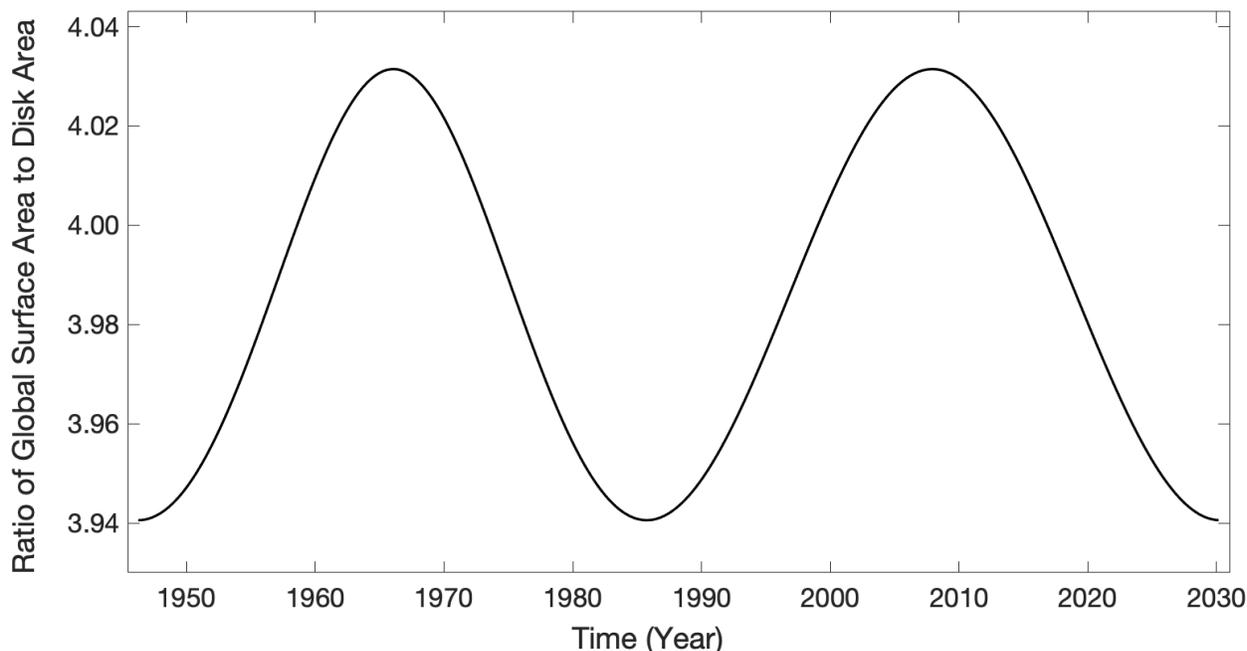

***Figure S2. Ratio of the global surface area to the solar-illuminated disk area.*** *The solar-illuminated disk varies over time with the changing sub-solar latitude.*

Figure S2 shows the ratio of global surface area to disk area for Uranus during 1946-2030, indicating a variation of ~2.3%, from ~3.94 at equinoxes to ~4.03 at solstices. Applying this ratio to the solar flux at Uranus (Panel C of Fig. S1) yields the global-average solar flux (Fig. S3). In Fig. S3, the solar longitude ($L_s$) is displayed to differentiate the NH seasons: spring (0-90°), summer (90-180°), autumn (180-270°), and winter (270-360°). Additionally, Uranus' perihelion and aphelion in its orbit around the Sun, corresponding to maximal and minimal solar flux, are marked. The global-average solar flux serves as the reference for computing the global-average absorbed solar power with the measured Bond albedo.

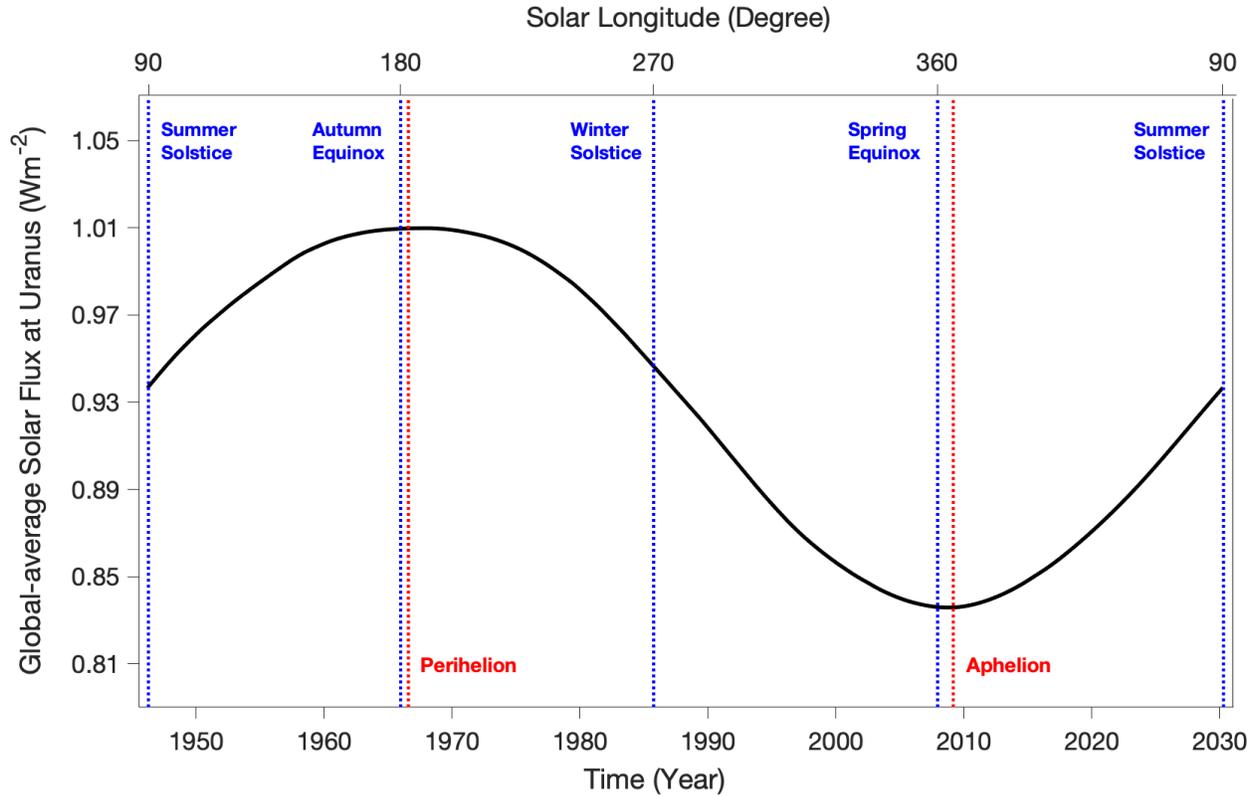

*Figure S3. Global-average solar flux at Uranus.* *The five blue vertical dashed lines indicate the summer solstices (1946 and 2030), autumn equinox (1966), winter solstice (1985), and spring equinox (2007) of the NH, respectively. The two red vertical dashed lines represent Uranus' perihelion (1966) and aphelion (2009) in its orbit around the Sun.*

We also examine Uranus' REB at the hemispheric scale in this study. We first construct the meridional distribution of solar flux during the period of 1946-2030 by considering orbital eccentricity, obliquity of rotational axis, and incidence angle of solar radiance (Fig. S4A). Then, the meridional distribution of solar flux is used to construct the hemispheric-average solar flux for the two hemispheres: the Northern Hemisphere (NH) and the Southern Hemisphere (SH). As shown in Fig. S4B, seasonal variations of solar flux are much stronger at the hemispheric scale than at the global scale, which is attributed to Uranus' extreme axial tilt (97.77°). The hemispheric-average solar flux is used to calculate the hemispheric-average absorbed solar power based on the hemispheric albedo. The Bond albedo at both global and hemispheric scales, along with its seasonal variations from 1946 to 2030, is discussed in the following section.

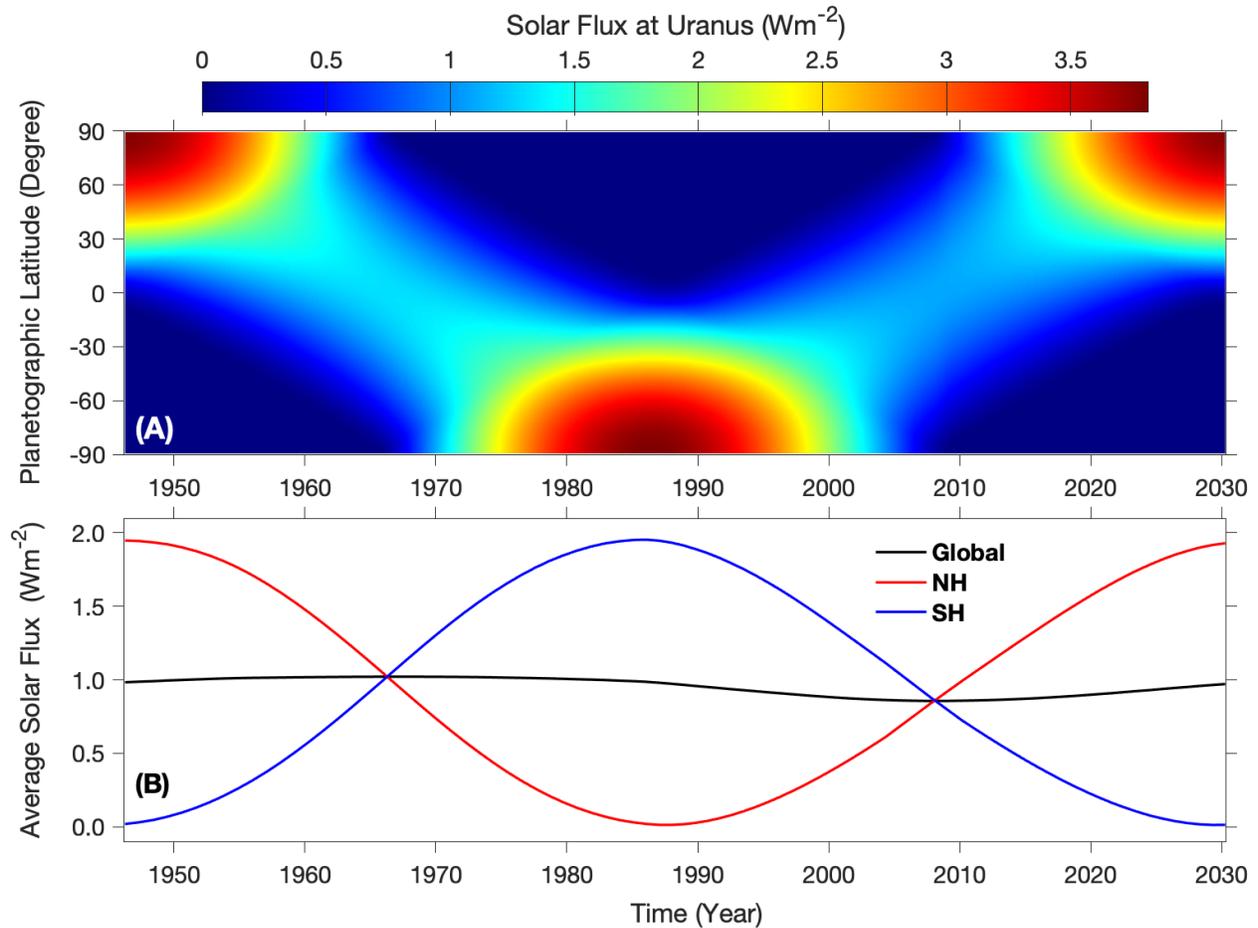

*Figure S4. Meridional distribution of solar flux at Uranus and hemispheric-average solar flux.*
*(A) Meridional distribution of solar flux at Uranus during the period of 1946-2030. (B) Hemispheric-average solar flux at Uranus. Global-average solar flux is also plotted in panel B for comparison between global and hemispheric averages.*

## S4. Bond Albedo and Absorbed Solar Power During the Period 1946–2030
### I. Seasonal Variations of Disk-Average Reflectivity

Uranus' geometric albedo exhibits seasonal variations. A recent study by Irwin et al. (2024), based on measurements from the Lowell Observatory (e.g., Lockwood and Jerzykiewicz, 2006; Lockwood, 2019) and the HST/WFC3, provides the seasonal variations of Uranus' disk-average reflectivity (I/F) from 1950 to 2022 (Fig. S5) at two wavelengths: blue (0.47 μm) and yellow (0.55 μm).

It is important to note that disk-average reflectivity does not represent global-average reflectivity. Uranus has a unique obliquity of 97.77°, which is close to 90°. As a result, only one hemisphere of Uranus is primarily illuminated by the Sun during the summer and winter solstices. During these times, the disk-average reflectivity represents hemispheric reflectivity rather than global reflectivity. The disk-average reflectivity discussed in this study pertains to the solar-illuminated disk, so disk-average reflectivity, rather than global-average albedo, should be used to determine the reflected and absorbed solar powers.

Figure S5 shows that reflectivity reaches the maxima around the summer and winter solstices and the minima around the spring and autumn equinoxes at both the blue and yellow wavelengths. These variations are probably caused by spatial variations in albedo (e.g., brighter polar regions compared to lower latitudes) (e.g., Lockwood, 1978; Karkoschka, 2001; Lockwood and Jerzykiewicz, 2006) and seasonal variations in albedo (e.g., Rages et al., 2004; Hammel and Lockwood, 2007). However, distinguishing between these two variations is challenging because the latitudinal coverage of the solar-illuminated area in full-disk albedo observations (Fig. S5) changes over time, intertwining spatial and temporal variations. In this study, we focus on the reflected and absorbed solar power from the solar-illuminated disk. Therefore, it is sufficient to examine the reflectivity of the solar-illuminated disk and its seasonal variations (Fig. S5), and we do not need to separate spatial and temporal variations.

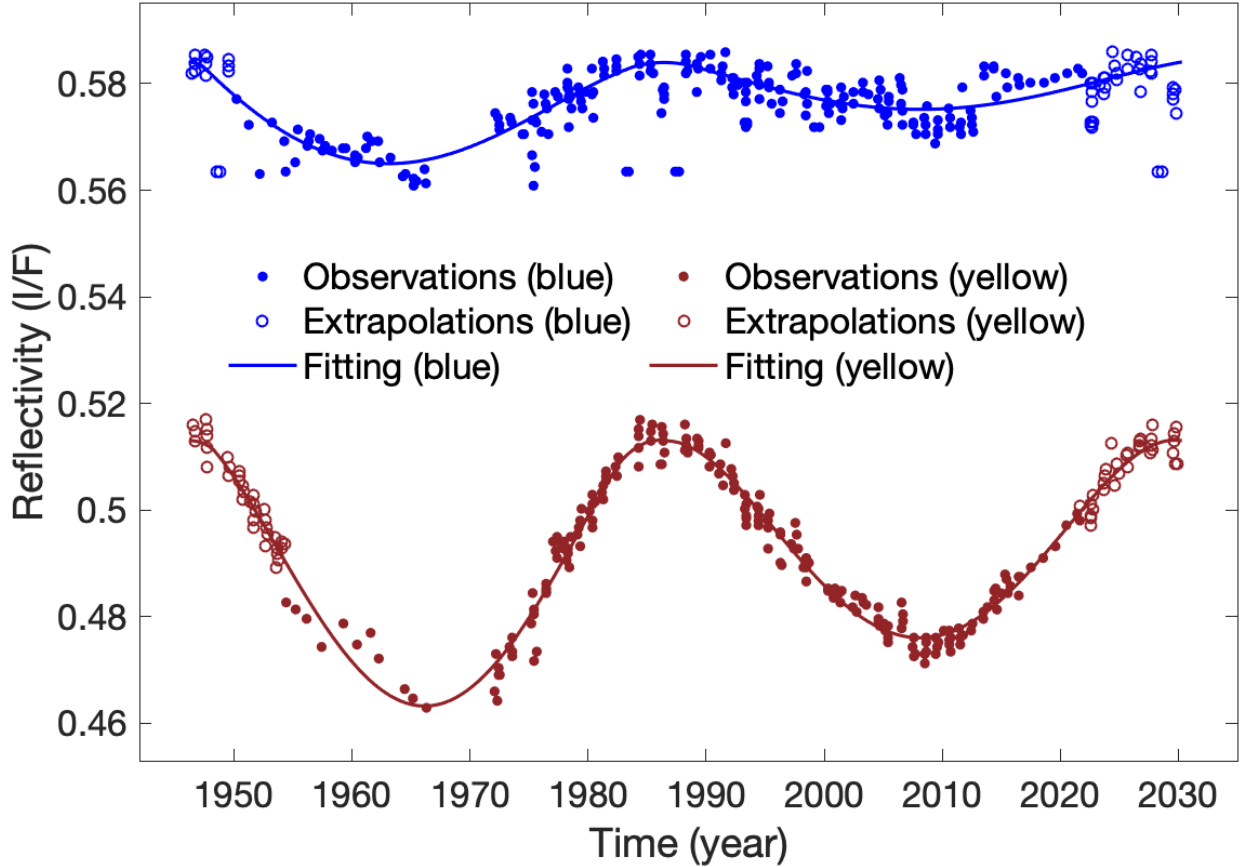

*Figure S5. Seasonal variations of disk-average reflectivity (I/F) at two wavelengths (blue at 0.47 μm and yellow at 0.55 μm). Solid dots represent observations, which circles stand for projected reflectivity for observational gaps. Solid lines show the fitting results (see discussion in S4).*

Since the temporal range of Fig. S5 (1950-2022) does not cover the full orbital period from 1946 to 2030, we first extend the measurements to the complete period. Figure S5 suggests that disk-average reflectivity follows a periodic pattern, reaching maxima at solstices and minima at equinoxes. To fill the observational gaps from 1946 to 1950, we mirror-project the observations from the period 1966-1985 (i.e., from the autumn equinox to the winter solstice in the NH) onto the period 1946-1966 (i.e., from the summer solstice to the autumn equinox in the NH) by assuming a symmetry around the autumn equinox in 1966. Similarly, we apply this mirror projection to extend the dataset from 2022 to 2030.

Next, we fit polynomial functions to the time series of disk-average reflectivity over the period 1946-2030, as shown in Fig. S5. These fitting curves represent the seasonal variations of Uranus' reflectivity at the blue and yellow wavelengths. We further interpolate and extrapolate these seasonal variations across the wavelength range of 0.12-4.16 μm. This wavelength range is selected based on two considerations: (1) the spectral solar irradiance within this range accounts for more than 99% of total solar power, and (2) observational data on Uranus' disk-average geometric albedo are available in this wavelength range.

**II. Disk-Average Geometric Albedo**

Figure S6 presents Uranus' geometric albedo over the wavelength range of 0.12-4.16 μm, derived from various observations and studies using different missions and instruments. These include the Voyager Ultraviolet Spectrometer (Voyager/UVS) at 0.12-0.17 μm (Yelle et al., 1989), the International Ultraviolet Explorer (IUE) at 0.17-0.30 μm (Wagener et al., 1986; Caldwell et al., 1988; Cochran et al., 1990), the European Southern Observatory (ESO) at 0.3-1.05 μm (Karkoschka, 1998), the NASA Infrared Telescope Facility (IRTF) at 1.05-2.50 μm (Neff et al., 1985), and the European Infrared Space Observatory Photopolarimeter (ISOPHOT) at 2.50-4.16 μm (Encrenaz et al., 2000). These observations were conducted at different times and wavelengths. Among them, high-quality, high-spectral-resolution spectra at 0.30-1.05 μm were recorded by ESO in 1995 at a phase angle of 0.7° (Karkoschka, 1998), providing a reliable approximation of geometric albedo in this wavelength range. Observations at other wavelengths were also recorded at phase angles close to 0°, which can be used to represent the geometric albedo.

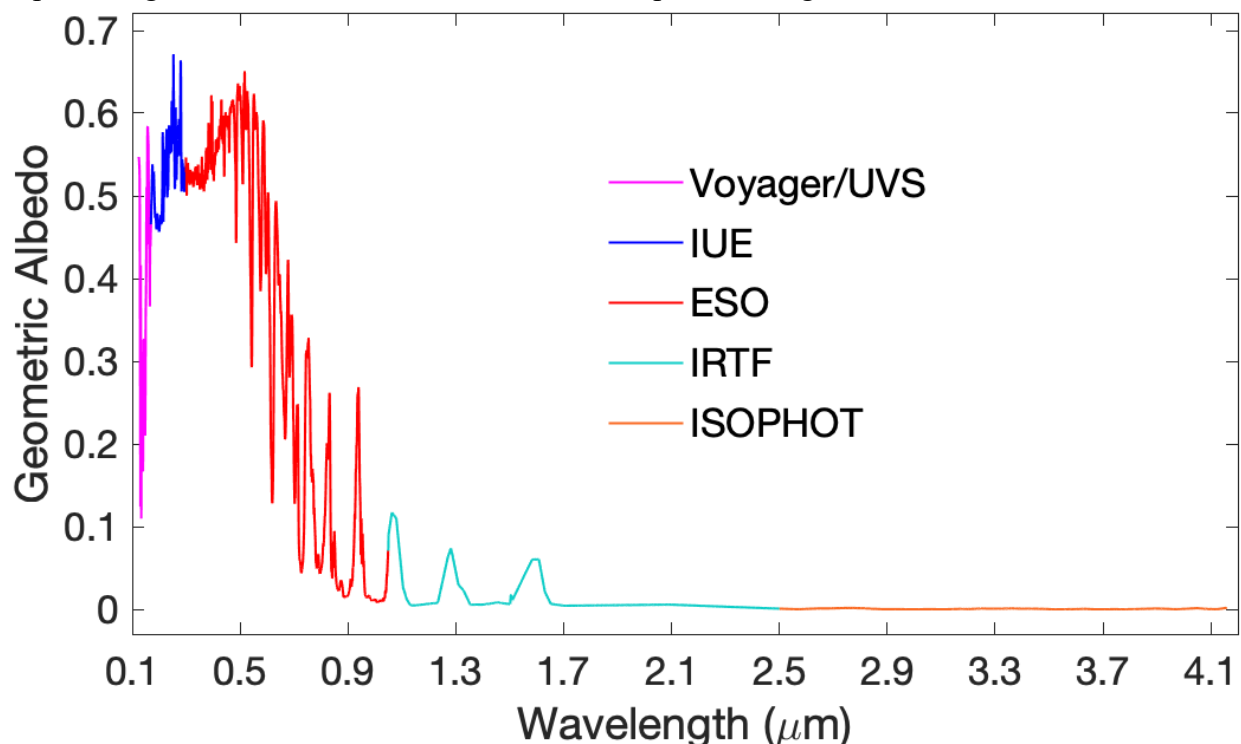

*Figure S6. Geometric albedo in the wavelength range of 0.12-4.16 μm. Observations at different wavelengths from various missions and instruments are used to construct the geometric albedo (see discussion in S4).*

The solar spectral irradiance within the ESO range (0.3-1.05 μm) contributes significantly to the total solar power. To ensure consistency, we use the polynomial fitting functions of seasonal cycles at blue and yellow wavelengths (Fig. S5) to project observations from times different from 1995 onto 1995. This adjustment allows for a uniform spectral representation of geometric albedo in 1995, as shown in Fig. S6. The spectral albedo is then used to compute the wavelength-average geometric albedo, weighted by solar spectral irradiance (Li et al., 2023), yielding a value of 0.263.

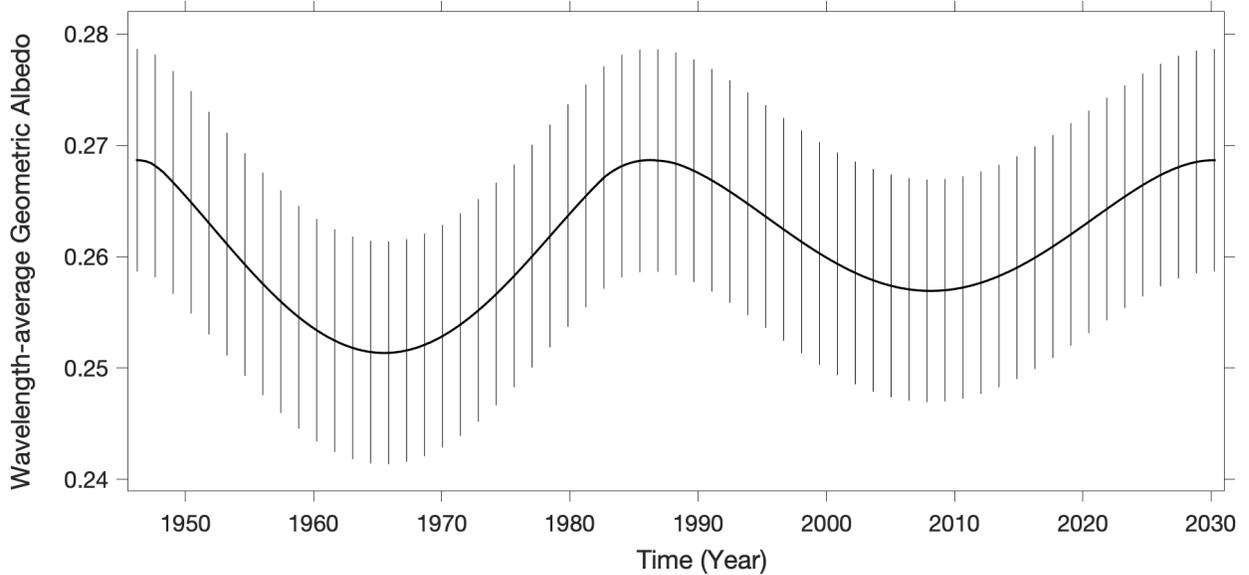

*Figure S7. Seasonal variations of wavelength-average geometric albedo. The seasonal variations at the blue and yellow wavelengths (Fig. S5), the spectra of geometric albedo (Fig. S6), and the solar spectral irradiance (Li et al., 2023) are used to construct the wavelength-average geometric albedo. The vertical lines represent uncertainties.*

Next, we assess the uncertainty in the estimated wavelength-average geometric albedo. There are two primary sources of uncertainty: (1) the uncertainty of measurements of geometric albedos at different wavelength ranges; and (2) the uncertainty related to projecting the measurements of geometric albedo from different times onto 1995 (i.e., the ESO observational time). The ESO measurements of geometric albedo at 0.3-1.05 μm have an uncertainty of ~ 4% (Karkoschka, 1998). Most observations at other wavelengths (Fig. S6) do not provide explicit uncertainty estimates. Here, we assume they have the same 4% uncertainty as that of the ESO measurements. Given that the ESO wavelength range dominates solar power contribution, this assumption does not significantly impact the estimated wavelength-average geometric albedo. To estimate the uncertainty of projecting measurements of geometric albedo from different times onto 1995, we use the ratio between the standard deviation and the mean value of reflectivity at blue and yellow wavelengths (Fig. S5): the ratios at 0.47 μm (blue) and 0.55 μm (yellow) are linearly interpolated and extrapolated to other wavelengths to estimate the uncertainties over there.

Combining the two uncertainties, we can estimate the total uncertainty in the wavelength-average geometric albedo. Then we have the wavelength-average geometric albedo with uncertainty as $0.263 \pm 0.010$. This result aligns well with previous measurements, such as $0.264 \pm 0.008$ (Lockwood et al., 1983) and $0.27 \pm 0.02$ (Neff et al., 1985), despite these previous measurements are based observations with the lower quality and limited wavelength coverage. Since the wavelength-average geometric albedo computation is based on the 1995 spectrum (Fig. S6), we use the interpolated/extrapolated seasonal variations at blue and yellow wavelengths (Fig. S5) to estimate the wavelength-average geometric albedo for other years. The seasonal variations of wavelength-average geometric albedo over the period 1946-2030 are shown in Fig. S7.

**III. Phase Function and Phase Integral**

As introduced in the methodology section (S1), Bond albedo can be computed by multiplying the geometric albedo by the phase integral. The best study of Uranus' phase function in previous investigations is based on Voyager observations (Pearl et al., 1990) because the flyby observations by the Voyager spacecraft provided much better coverage of phase angle compared to these observations recorded by ground-based telescopes and Earth-orbiting observatories. A recent study (Wenkert et al., 2022) recalibrated the reflectivity data recorded by the Voyager Imaging Science Subsystem (ISS) (Smith et al., 1977). Here, we use the recalibrated reflected data recorded with a clear filter (0.30-0.64 μm). Additionally, ESO observations (Karkoschka, 1998) provide measurements of reflectivity at a 0.7° phase angle, which is close to the geometric albedo. Using the transmission data of the Voyager/ISS clear filter, we can convert the spectral-revolved albedo recorded by ESO to the clear-filter data at the 0.7° phase angle. Finally, the Voyager Infrared Interferometer Spectrometer and Radiometer (IRIS) (Hanel et al., 1980) recorded Uranus' reflectivity at different phase angles across a wavelength range of 0.3-2.0 μm. These measurements are combined to examine the phase function and phase integral of Uranus' disk-average reflectivity.

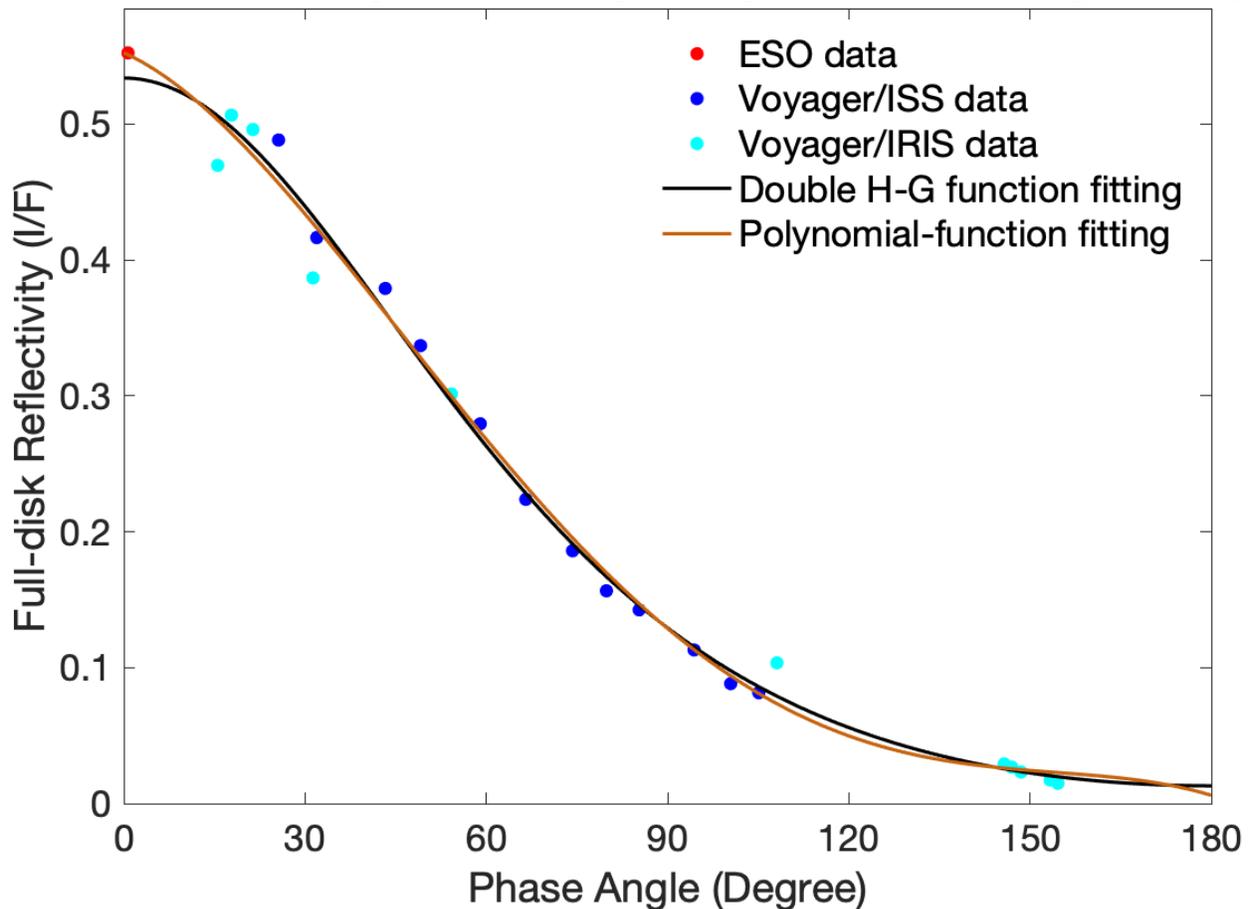

*Figure S8. Fitting the phase function of Uranus' full-disk reflectivity. A fourth-order polynomial function (brown line) and the double H-G function (black line) are used to fit the combined data from the ESO observations and the Voyager ISS/IRIS observations.*

Figure S8 presents the combined measurements, which offer the best coverage of phase angles among all existing analyses. However, even with the combined measurements, there are still gaps in phase angle coverage. Our previous studies of the Bond albedo of Jupiter and Saturn (Li et al., 2018; Wang et al., 2024) suggest that polynomial functions using the least-squares technique (Bevington and Robinson, 2003) can efficiently fill the observational gaps in phase angle for the giant planets. For Uranus, we tried different polynomial functions and found that a fourth-order polynomial function fits the observed data best with the smallest fitting residual. We also tried physically-based functions, such as the double Henyey-Greenstein (H-G) function (Henyey and Greenstein, 1941; Hapke, 2002), which also fit the observed data well. Figure S8 presents the fittings using both the polynomial and H-G functions. The fitting comparison suggests that they are generally consistent, except for small phase angles near 0°, where the polynomial function fits the data better. Therefore, the polynomial-function fitting is used in our analysis of Uranus' phase integral, yielding a value of 1.30.

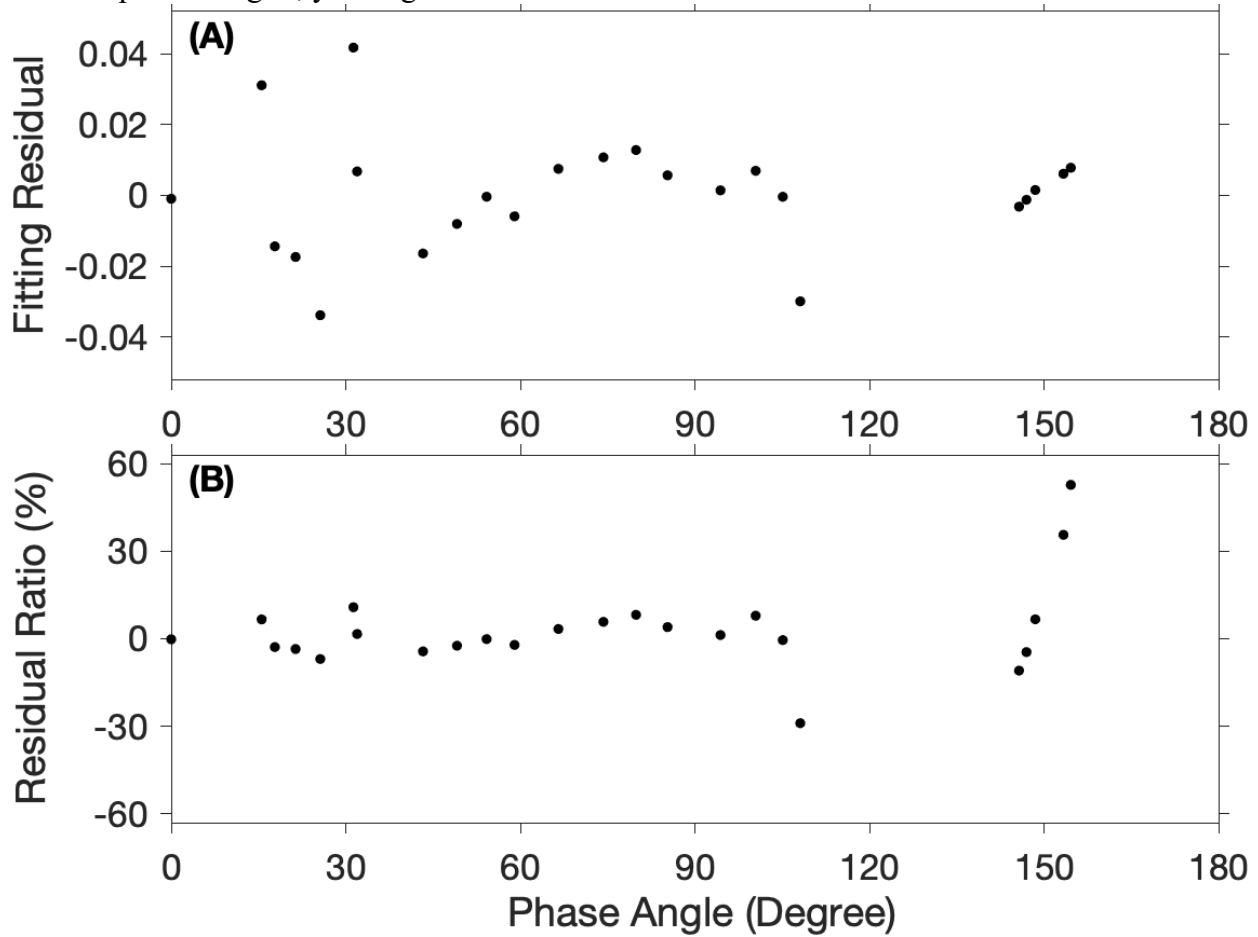

*Figure S9. Fitting residuals for the fits shown in Fig. S8.* *(A) Fitting residuals. (B) Ratios between fitting residuals and the observed data. Fitting residuals are defined as the fitting results minus the observational data.*

We also examined the fitting residuals (i.e., the difference between the fitted values and measurements) and the ratio of the residuals to the measurements (Fig. S9), which can be used to assess the fitting quality. Fig. S9B suggests that the residual ratios are less than 15%, except for the measurement at a phase angle around 108° and measurements with phase angles greater than 150°. These large residual ratios are partly caused by the relatively small reflectivity values in those regions (see Fig. S8). The fitting residuals are also used to estimate the uncertainty of phase integral, which is related to filling observational gaps in phase angle by fitting. The uncertainty at each observed phase angle is represented by the corresponding fitting residual, and these estimates are then interpolated or extrapolated to phase angles without observations. Finally, the estimated uncertainties are substituted into the phase integral formula (e.g., Li et al., 2023) using the propagation rule of addition (Bevington and Robinson, 2003). Our analysis suggests that the phase integral has an uncertainty of ~0.03. Therefore, the phase integral can be expressed as 1.30±0.03.

The best estimate of the phase integral from Pearl et al. (1990) is 1.40±0.14. The difference between our analysis and the previous estimate is smaller than the relatively large uncertainty in the previous study. The reflectivity at a 0° phase angle plays a critical role in the phase integral (e.g., Conrath et al., 1989; Li et al., 2023). Since the Voyager spacecraft did not conduct full-disk observations near the 0° phase angle, the reflectivity around this phase angle was not included in the previous analysis by Pearl et al. (1990), which may contribute to the relatively large uncertainty in their study.

### IV. Disk-Average Bond Albedo

Based on the wavelength-average geometric albedo and phase integral, we calculate the Bond albedo as 0.263×1.30 = 0.342. Assuming that the uncertainties in the measurements of the wavelength-average geometric albedo and phase integral are independent, we apply error propagation for multiplication (Bevington and Robinson, 2003) to estimate the uncertainty in the Bond albedo, yielding a value of 0.017. Therefore, the Bond albedo of Uranus in 1995 is 0.342±0.015. It is possible that the seasonal variations in geometric albedo (i.e., reflectivity at 0° phase angle) (Fig. S5) represent the seasonal variations in reflectivity at other phase angles. Then the phase integral does not vary over time. Therefore, the seasonal variations shown in Fig. S7 and the phase integral based on Fig. S8 can be used to estimate the seasonal variations of Uranus' Bond albedo, as shown in Fig. S10.

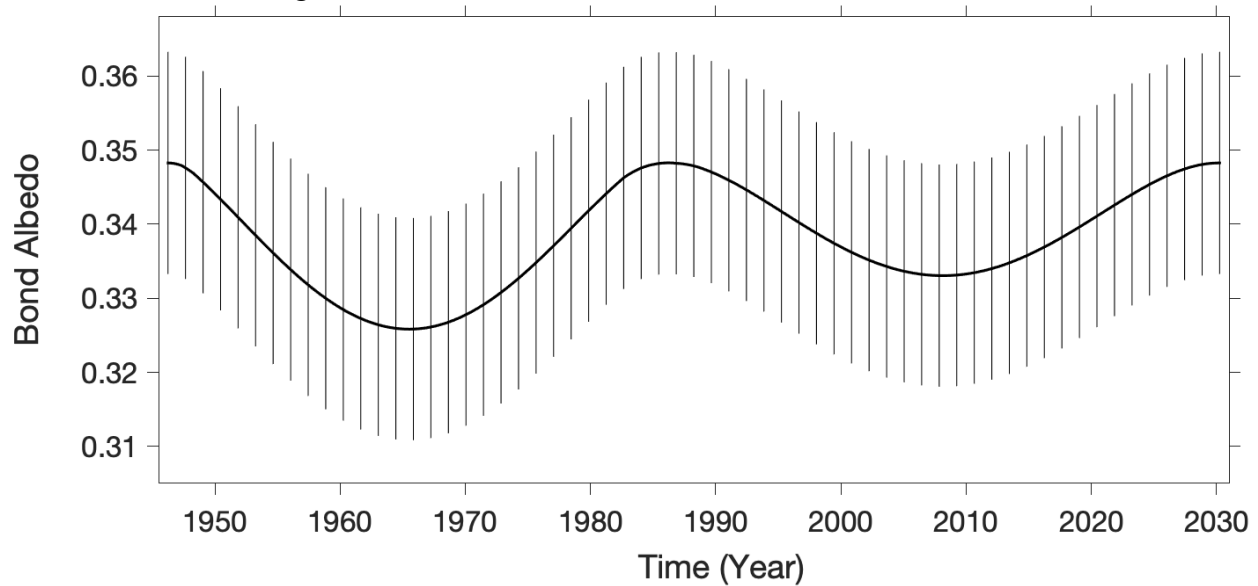

*Figure S10. Seasonal variations of disk-average Bond albedo.* The seasonal variations of Uranus' Bond albedo are based on the seasonal variations of geometric albedo (Fig. S7) and the phase function (Fig. S8). The phase function and the related phase integral are assumed to be constant over time. The vertical lines represent uncertainties.

### V. Hemispheric Albedo

In this study, we also investigate the REB at the hemispheric scale. Constructing the albedo of the solar-illuminated hemisphere requires regional observations with good coverage of view geometry. However, the currently available observations are insufficient to measure such an albedo. Fortunately, the unique obliquity of Uranus' rotational axis allows for the estimation of hemispheric albedo from the disk-average Bond albedo. Uranus' obliquity (97.77°) enables the use of disk-average Bond albedos (Fig. S10) around a hemisphere's summer solstice to represent the albedo of that hemisphere. In other words, the disk-average Bond albedo around the NH summer solstice can be used to approximate the NH albedo, while the disk-average Bond albedo around the NH winter solstice (i.e., the SH summer solstice) can be used to represent the SH albedo.

A recent study (Wang et al., 2024) suggests that bright latitudinal bands created by giant convective storms do not significantly affect Saturn's Bond albedo at the hemispheric scale, meaning Saturn's NH and SH albedos are essentially the same. Here, we assume that Uranus' albedo does not vary significantly between the two hemispheres, so the disk-average Bond albedo during the spring/autumn equinox can represent the albedo in each hemisphere. Thus, the disk-average Bond albedo during the periods of 1946-1966 (from summer solstice to spring equinox of the NH) and 2007-2030 (from spring equinox to summer solstice of the NH), which is presented in Fig. S10, can represent the average albedo of the solar-illuminated areas in the NH during the same periods. Likewise, the disk-average Bond albedo from 1966-2007 can approximate the average albedo of the solar-illuminated areas in the SH during the same periods.

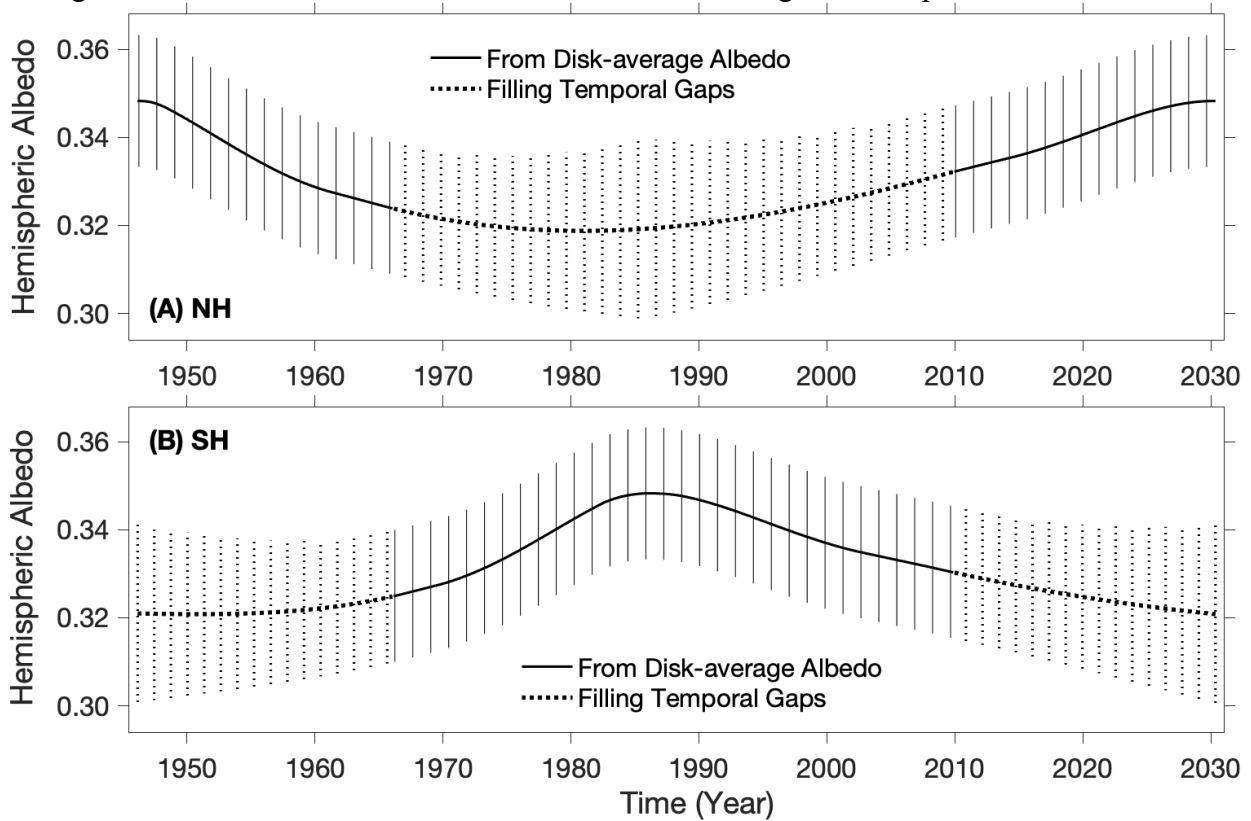

*Figure S11. Seasonal variations of Bond albedo at the hemispheric scale. The seasonal variations of Uranus' Bond albedo at the hemispheric scale are based on the seasonal variations of disk-average Bond albedo (Fig. S10) for certain seasons (solid lines), which are interpolated/extrapolated to other seasons (dashed lines) (see discussion in S4). The vertical lines stand for uncertainties.*

Figure S11 presents the average albedo for the solar-illuminated areas in each hemisphere. For the time periods represented by solid lines, we use the disk-averaged albedo (Fig. S10) to approximate the hemispheric albedo, as discussed above. Polynomial functions are then used to fit these solid lines and to fill the temporal gaps (dashed lines in Fig. S11). Figure S11 shows that the fitted hemispheric albedo in the temporal gaps has relatively low values. During these temporal gaps, which occur around the winter solstice of each hemisphere, the solar-illuminated region is primarily confined to low latitudes, where reflectivity is generally low (e.g., Lockwood, 1978; Karkoschka, 2001; Lockwood and Jerzykiewicz, 2006; Irwin et al., 2024). This is consistent with the fitting results in Fig. S11.

After filling the gaps, we can determine the average albedo for the solar-illuminated areas in each hemisphere during the complete orbital period (1946-2030). To estimate the uncertainty in the albedo for the NH and SH hemispheres, we use the uncertainty in the disk-average Bond albedo (Fig. S10) as the uncertainty in the hemispheric albedo for the periods extracted from the disk-average albedo (i.e., periods with solid lines in Fig. S11). In principle, the uncertainty of the filling hemispheric albedo in the temporal gaps (i.e., periods with dashed lines in Fig. S11) should be larger than the uncertainty during the extracted periods. The uncertainty in the temporal gaps should increase as the distance from the extracted periods increases. For example, we expect the uncertainty of the NH filling albedo to be greatest around the year 1986-87, which are in the middle of the NH temporal gaps (1966–2007). Similarly, we expect the uncertainty in the SH filling albedo to be highest in the years 1946 and 2030. We estimate the maximal uncertainty by adding the standard deviation of the albedo during the extracted period to the uncertainty of the hemispheric albedo for the extracted period. Subsequently, we linearly interpolate the uncertainty of hemispheric albedo from the extracted periods to the years with maximal uncertainty.

Figure S11 suggests that the uncertainties of the filling albedo in the temporal gaps are relatively large, comparable to the magnitudes of seasonal variations in hemispheric albedo and larger than the temporal variations of the filling lines themselves. More importantly, these temporal gaps correspond to the years around the winter solstice in each hemisphere when the incident solar flux is at a minimum for that hemisphere (see Fig. S4). Therefore, the absorbed solar power is extremely small for the hemisphere during winter solstice (see Fig. S12 in the next section), regardless of the value of hemispheric albedo. In other words, the large uncertainties for filling the temporal gaps do not significantly affect our subsequent discussion of the hemispheric REB.

**VI. Absorbed Solar Power at Global and Hemispheric Scales**

Based on the discussions of solar flux at Uranus and Uranus' albedo at global and hemispheric scales, we can determine the absorbed solar power of the giant planet. This study focuses on the global and hemispheric averages of REB. Therefore, the absorbed solar power and emitted thermal power are averaged over the global and hemispheric areas, with units of watts per unit area (W/m²). To simplify the discussion, we refer to the absorbed and emitted powers per unit area (in W/m²) as absorbed and emitted powers, respectively. Figure S12 shows the global and hemispheric absorbed solar powers by Uranus during the complete orbital period from 1946 to 2030. The uncertainty in absorbed solar power is determined by the uncertainty in the albedo measurements. Figure S12 suggests that seasonal variations in absorbed solar power are much stronger on the hemispheric scale than on the global scale. The comparison of absorbed solar power between global and hemispheric scales shows a similar pattern to that of solar flux at Uranus (see Fig. S4). The seasonal variations are much stronger in solar flux (~17% at the global scale and ~100% at the hemispheric scale) than in albedo (~6% for both global and hemispheric scales), so the seasonal variations in Uranus' absorbed solar power are primarily determined by the seasonally varying solar flux.

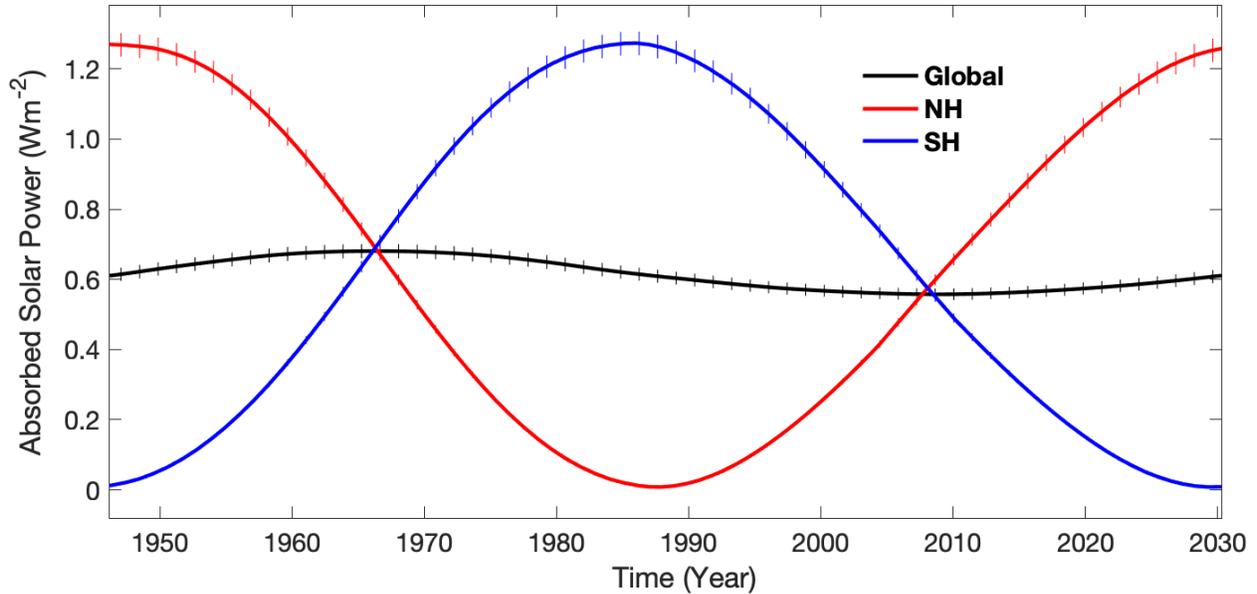

*Figure S12. Absorbed solar power at the global and hemispheric scales. The solar flux at the global and hemispheric scales (Fig. S4) is combined with the Bond albedo at the global and hemispheric scales (Figs. S10 and S11) to compute the absorbed solar power at the global and hemispheric scales. The vertical lines indicate uncertainties.*

## S5. Emitted Thermal Power During the Period of 1946-2030

After discussing Uranus' absorbed solar energy, we now turn to another component of the REB: emitted thermal energy. The emitted thermal power from Uranus has been investigated in many previous studies (e.g., Fazio et al., 1976; Loewenstein et al., 1977; Courtin et al., 1978; Stier et al., 1978; Hildebrand et al., 1985; Orton, 1985; Neff et al., 1985; Moseley et al., 1985; Pearl et al., 1990). Most of these earlier measurements were based on observations with significant limitations in wavelength and emission-angle coverage, leading to large uncertainties. The most systematic analysis of Uranus' emitted power comes from a study by Pearl et al. (1990), which is based on thermal observations of Uranus by the Voyager spacecraft. The results from this study are used in our analysis of Uranus' REB.

We now examine the seasonal variations in Uranus' emitted power. The thermal structure of Uranus' lower stratospheric and upper troposphere, which are primarily responsible for the emitted power, does not exhibit significant seasonal changes, mainly because its radiative time constant is longer than its orbital period (Conrath et al., 1990). As a result, the emitted thermal power is also expected to have insignificant seasonal variations. Observations of brightness temperature support this, indicating that the lower stratosphere and upper troposphere do not show significant seasonal changes (e.g., Orton et al., 2015; Roman et al., 2020). Numerical simulations also suggest that Uranus' atmosphere and its emitted power do not undergo significant seasonal variations (e.g., Wallace, 1983; Friedson and Ingersoll, 1987; Conrath et al., 1990; Bezard and Gautier, 1986; Bezard, 1990; Milcarek et al., 2024).

Thermal radiance, or brightness temperature, at far-infrared wavelengths has been used in some early studies to estimate the emitted power and effective temperature of Uranus (e.g., Fazio et al., 1976; Stier et al., 1978; Hildebrand et al., 1985). Recent studies of emitted powers from other planets (Li et al., 2010, 2012; Creecy et al., 2022) suggest that such estimates have large uncertainties, as they do not account for the dependence of outgoing thermal radiance on wavelength and emission angle. On the other hand, measurements of Jupiter's emitted power based on Cassini observations reveal that the outgoing thermal radiance is correlated across different wavelength ranges and emission angles (Li et al., 2012). Therefore, we expect that temporal variations in outgoing thermal radiance at specific wavelengths may provide insights into the temporal variations of total emitted power, even though thermal radiance at limited wavelengths cannot fully represent the total emitted power.

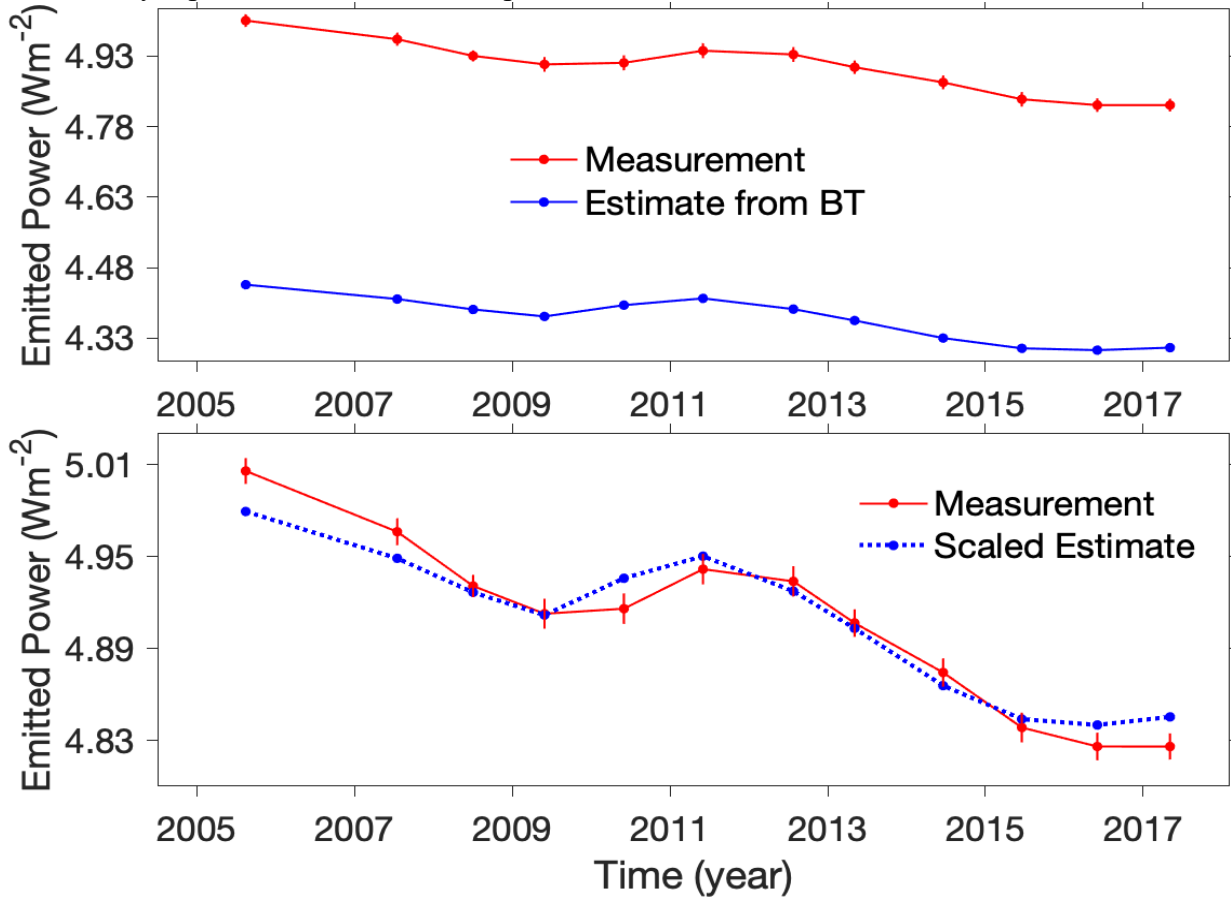

*Figure S13. Comparison of Saturn's emitted power between direct measurements and estimates based on brightness temperature. (A) Comparison of global-average emitted power between measurements and estimates from brightness temperature. The measurements are from a recent study (Wang et al., 2024), and the estimates are obtained by applying the Stefan-Boltzmann law to the global-average brightness temperature. (B) Comparison of global-average emitted power between measurements and scaled estimates. The estimates of emitted power from brightness temperature shown in panel (A) are scaled by a factor corresponding to the ratio of time-mean values between the estimated and measured emitted power (see discussion in S5).*

Here, we use long-term thermal observations of Saturn recorded by the Cassini Composite Infrared Spectrometer (CIRS) (Flasar et al., 2004) to explore the possibility of using outgoing thermal radiance at limited wavelengths to examine temporal variations in emitted power. Previous studies (Allison and Travis, 1986; Li et al., 2010) suggest that thermal radiance from pressure levels around 300 mbar primarily contributes to Saturn's total emitted power. The introductory paper for Cassini CIRS indicates that pressure levels around 300 mbar are predominantly sensed by spectra at a wavenumber of 290 cm$^{-1}$ (~34.5 μm). We select thermal radiance in the wavenumber range of 260-320 cm$^{-1}$ (~31.2-38.5 μm) to examine the seasonal variations of Saturn's emitted thermal radiance during the Cassini epoch (2004-2017). A recent study (Wang et al., 2024) provided measurements of Saturn's emitted power during the Cassini epoch, considering the dependence of outgoing thermal radiance on wavelength and emission angle. Comparing the seasonal variations between the thermal radiance around 290 cm$^{-1}$ and the emitted power allows us to assess any correlation between them.

We first organize all CIRS spectra wavenumber range of 260-320 cm$^{-1}$ into a two-dimensional matrix (year × latitude) of thermal radiance during the Cassini epoch. We then convert the radiance into brightness temperature (BT) by assuming blackbody radiation. The BT at different latitudes is averaged meridionally to obtain the global-average BT during the Cassini epoch. Applying the Stefan-Boltzmann law to the global-average BT, we estimate the global-average emitted power during the Cassini epoch. Panel A of Fig. S13 compares the estimated emitted power with the measured emitted power (Wang et al., 2024), suggesting that the estimated emitted power from BT significantly underestimates Saturn's emitted power. However, the comparison also indicates that both the estimated and measured emitted powers show similar temporal variations. To better compare the temporal variations between the two emitted powers, we scale the estimated emitted power by a factor corresponding to the ratio of the time-mean values between the estimated and measured emitted powers. This ensures that the estimated and measured emitted powers have the same time-mean value. Panel B of Fig. S13 compares the scaled estimate of the emitted power from BT with the measured emitted power, demonstrating similar temporal variations between them. In other words, the scaled BT-based emitted power can be used to approximate Saturn's emitted power and its temporal variations.

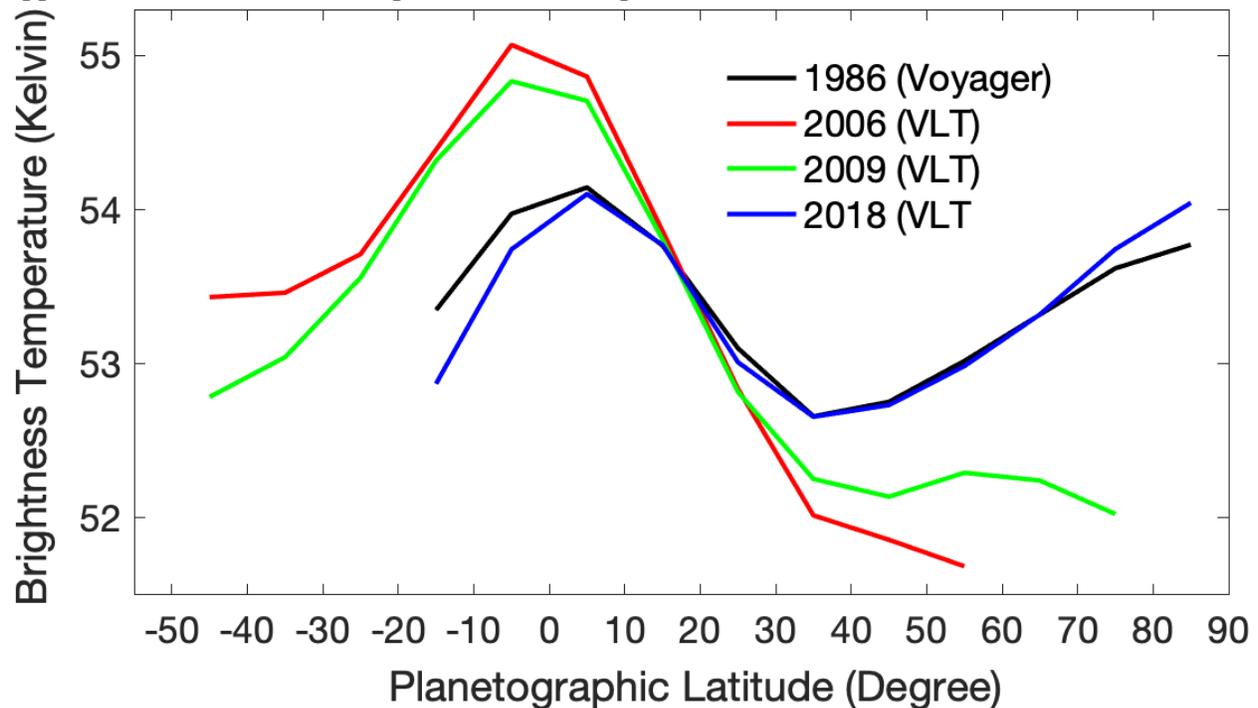

*Figure S14. Meridional profiles of brightness temperature.* *Data are from two previous studies (Orton et al., 2015; Roman et al., 2020). The profiles in 2006 and 2018 were observed at 18.7 μm, and the profile in was obtained at 19.5. The observations for the profiles in the three years (2006, 2009, and 2018) were recorded by the VISIR instrument at the Very Large Telescope (VLT). The Voyager profile in 1986 was generated by forward modeling the temperatures derived from the*

*Voyager/IRIS spectra (Orton et al., 2015) to the same viewing conditions as the 2018 data (Roman et al., 2020).*

For Uranus, observational studies of seasonal variations in emitted power are lacking due to the unavailability of high-quality continuous observations with good coverage of wavelength and viewing geometry. Based on the Cassini tests of Saturn's emitted power discussed above, we use BT observations of Uranus to estimate the seasonal variations of this ice giant. Thermal radiance from pressure levels around 300-400 mbar primarily contributes to Uranus' emitted power (Allison and Travis, 1986). The thermal radiance at wavelengths of 18.7 μm and 19.5 μm predominantly originates from pressure levels around 300 mbar in Uranus' atmosphere (Orton et al., 2015). Therefore, BT observations at 18.7 μm and 19.5 μm can be used to approximate Uranus' emitted power and its seasonal variations. The Cassini tests of Saturn's emitted power suggest that we need meridional profiles and global averages of BT to approximate emitted power. However, observations of meridional profiles of Uranus' BT at 18.7 μm and 19.5 μm are limited. Recent studies (Orton et al., 2015; Roman et al., 2020) provide meridional profiles of BT for four years (1986, 2006, 2009, and 2018), which are shown in Fig. S14.

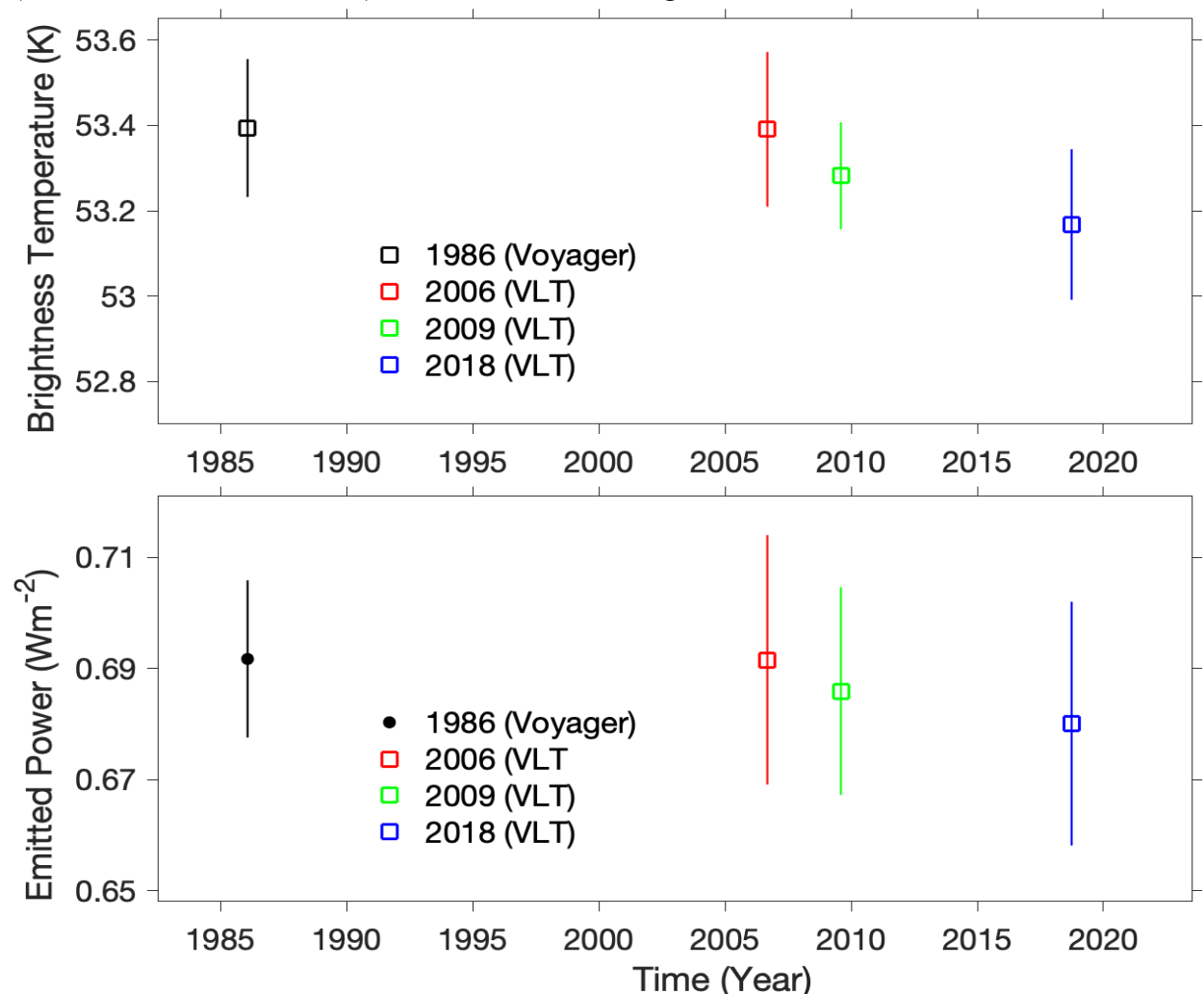

***Figure S15. Global-average brightness temperature and the corresponding estimated emitted power.*** *(A) Global-average brightness temperature based on Fig. S14. (B) Estimated emitted power derived from the global-average brightness temperature. In panel (B), the Voyager*

*measurement of global-average emitted power, shown as a dot, is from a previous study (Pearl et al., 1990). The Voyager measurement serves as a reference to estimate the emitted powers based on the brightness temperature for other years, which are shown as squares. Vertical lines represent uncertainties.*

The latitude coverage of the BT meridional profiles shown in Fig. S14 is incomplete. To estimate the global-average BT, we extrapolate BT from the observed latitudes to the unobserved latitudes by assuming that BT remains constant from the farthest observed latitude to the unobserved latitudes. The resulting global-average BT for the four years is shown in Fig. S15. The global-average BT values are 53.39±0.16 K, 53.39±0.18 K, 53.28±0.13 K, and 53.17±0.18 K for 1986, 2006, 2009, and 2018, respectively. We use the standard deviation of the meridional variations of the BT in the observed latitudes to represent the uncertainty in the extrapolated BT at the unobserved latitudes and then apply the error propagation rule (Bevington and Robinson, 2003) to estimate the uncertainty in the global-average BT.

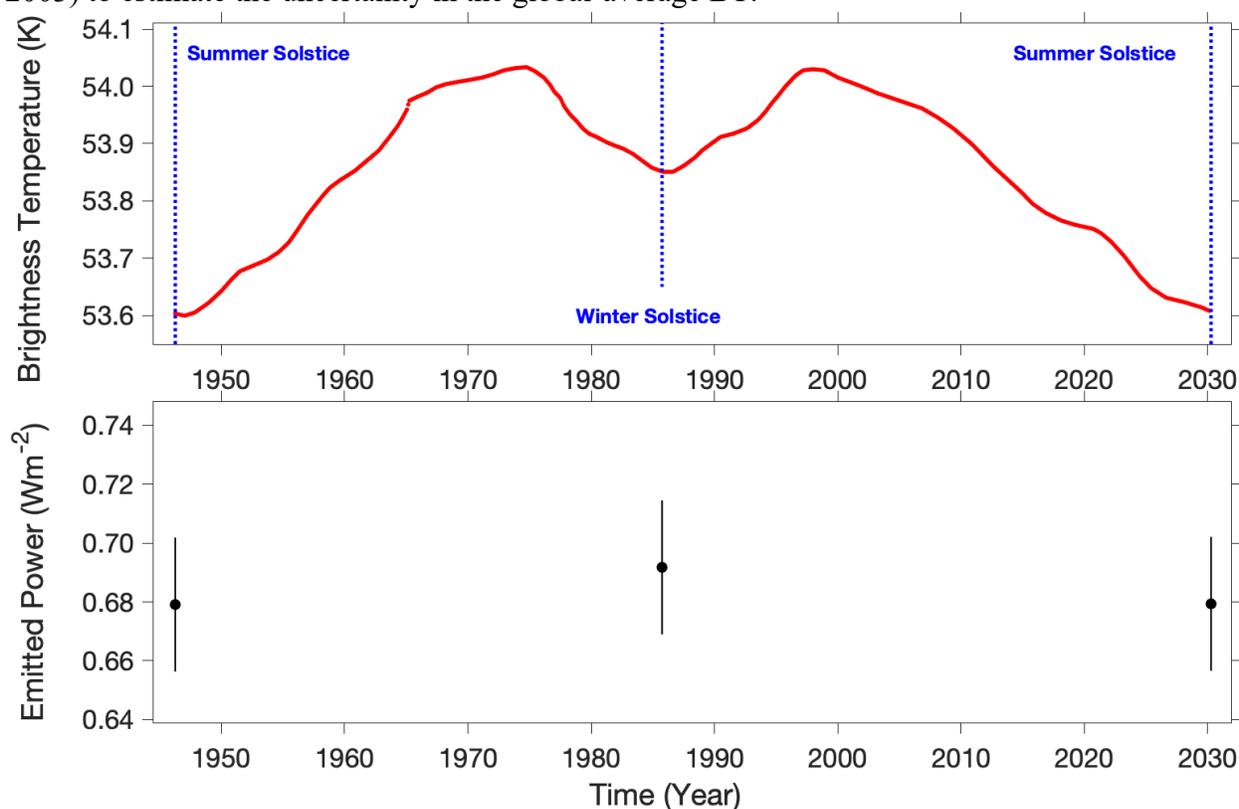

*Figure S16. Seasonal variations of disk-average brightness temperature and estimated emitted power. (A) Disk-average brightness temperature constructed in a previous study (Roman et al., 2020). The blue vertical dashed lines indicate the summer and winter solstices of the NH. (B) Estimated emitted power obtained by applying the Stefan-Boltzmann law to the disk-average brightness temperature and referencing the Voyager measurements near the winter solstice. The disk-average brightness temperature at the summer solstice of each hemisphere is used to represent the hemispheric-average brightness temperature. Vertical lines represent uncertainties.*

Next, we use the global-average BT to estimate Uranus' global-average emitted power by following the same method used with Cassini observations of Saturn's emitted power (see Fig. S13). The 1986 observations of Uranus' BT are based on the Voyager/IRIS. The Voyager IRIS observations are also used to obtain the best measurement of Uranus' emitted power (0.692±0.014

Wm$^{-2}$) (Pearl et al., 1990). This measurement serves as a reference for estimating Uranus' emitted power in other years and seasonal variations. We first apply the Stefan-Boltzmann law to the global-average BT presented in panel A of Fig. S15 to estimate the global-average emitted powers in the four years (1986, 2006, 2009, and 2018). The ratio between the estimated emitted power in 1986 and the Voyager measurement of Uranus' emitted power in the same year (0.692±0.014 Wm$^{-2}$) (Pearl et al., 1990) is used to scale the estimated emitted powers for the other three years (2006, 2009 and 2018). Panel B of Fig. S15 shows the Voyager-measured emitted power in 1986 and the scaled emitted powers in the other three years.

The uncertainties of the scaled emitted powers in the three years are estimated by considering three sources of uncertainty: (1) the uncertainty in the global-average BT caused by the incomplete latitudinal coverage of observations (Fig. S14); (2) the uncertainty in using global-average BT to estimate the global-average emitted power; and (3) the uncertainty in the Voyager measurement of emitted power in 1986 (Pearl et al., 1990). The first uncertainty is discussed and presented in panel A of Fig. S15. For the second uncertainty, we refer to the tests based on Cassini observations of Saturn's emitted power, as discussed in Fig. S13. We first compute the ratio of the difference between the scaled emitted power and the measured emitted power (see panel B of Fig. S13) to the measured emitted power. Then, we multiply this ratio by the scaled emitted power of Uranus (panel B of Fig. S15) to estimate the uncertainty. The third uncertainty, propagated from the error in the Voyager measurement, is taken from the previous study (Pearl et al., 1990). Panel B of Fig. S15 shows that the differences in emitted power among the four years (1986, 2006, 2009, and 2018) are smaller than the uncertainties, suggesting that the seasonal variations in Uranus' global-average emitted power are not statistically significant.

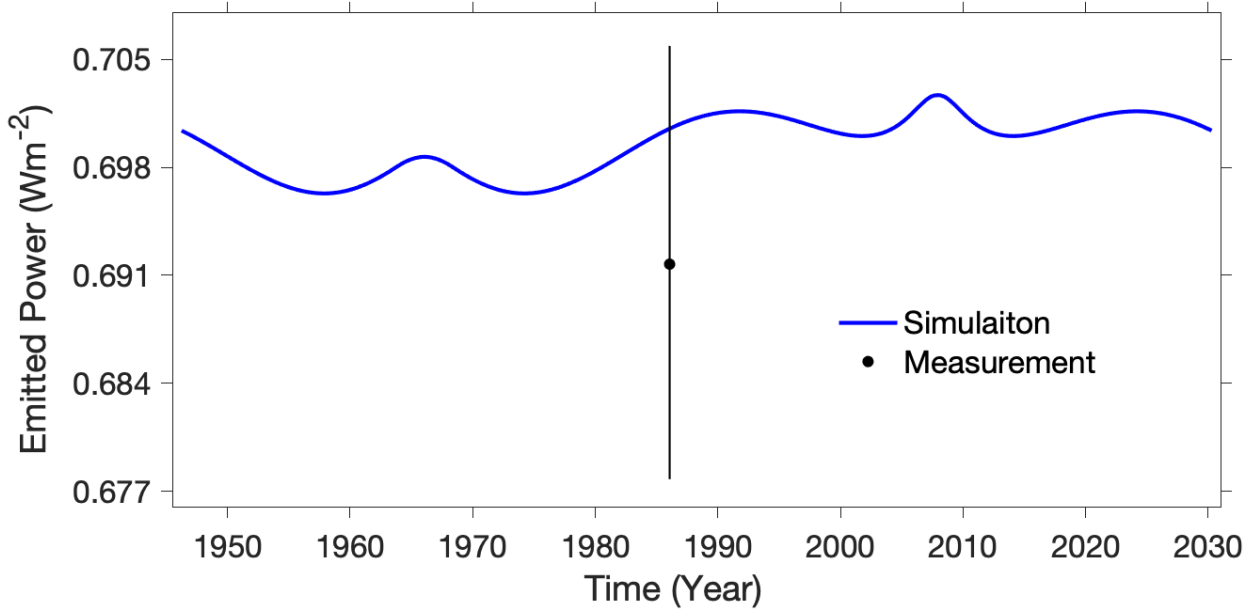

*Figure S17. Comparison of global-average emitted power between the Voyager measurement and simulation.* *The Voyager measurement comes from a previous study (Pearl et al., 1990) and the simulated emitted power is based in the simulations conducted in a previous numerical work (Wallace, 1983).*

Now, we discuss the hemispheric-average emitted power for our investigation of the REB at the hemispheric scale. In a recent study examining the BT of Uranus (Roman et al., 2020), the disk-average BT was constructed over a complete orbital period based on observations and a forward model. The disk-average BT is different from the global-average BT we discussed above because the former is an average of BT over the planetary disk of Uranus facing the observer (i.e., the observatories on Earth), while the latter is an average over all latitudes from pole to pole. The planetary disk of Uranus facing Earth has significantly different latitudinal coverage in different seasons due to its unique obliquity close to 90°. Fortunately, Uranus' planetary disk facing Earth during the solstices provides almost complete latitudinal coverage for one hemisphere, while the other hemisphere is largely not covered. Therefore, the disk-average BT can be used to approximate the hemispheric-average BT during the solstices. Panel A of Fig. S16 shows the seasonal variations of the disk-average BT constructed by Roman et al. (2020). The disk-average BT during the summer solstices of the NH (i.e., 1946 and 2030) can be used to approximate the NH-average BT, while the disk-average BT during the winter solstice of the NH (i.e., 1985) can be used to represent the SH-average BT. We then follow the same procedure for estimating emitted power from BT (Fig. S15) at the global scale to convert the hemispheric-average BT into hemispheric-average emitted power, as presented in panel B of Fig. S16. The comparison of estimated hemispheric emitted power between the NH (i.e., 1946 and 2030) and the SH (i.e., 1985) suggests that the difference of emitted power between the two hemispheres is smaller than the uncertainty in the estimated hemispheric emitted power. Therefore, we conclude that the hemispheric-average emitted power does not change from the NH to the SH. In other words, we assume that the hemispheric-average emitted power is the same as the global-average emitted power in our analysis of the REB at the hemispheric scale.

In addition to the examination based on BT observations, some numerical studies have been conducted to investigate the seasonal variations of Uranus' thermal structure (e.g., Wallace, 1983; Friedson and Ingersoll, 1987; Conrath et al., 1990; Bezard and Gautier, 1986; Bezard, 1990; Milcarek et al., 2024), which are closely related to the seasonal variations in emitted power. In particular, the simulated meridional variations of Uranus' effective temperature and emitted power are provided in the numerical work by Wallace (1983). Based on the meridional variations, we can derive the global-average emitted power. Furthermore, we project the simulated period to the orbital period from 1946 to 2030, which is the period investigated in this study. Figure S17 presents the global-average emitted power during the 1946-2030 period, based on the simulations conducted by Wallace (1983). The ratio of the standard deviation of the seasonal variations to the annual mean for the simulated global-average emitted power is ~0.3%, suggesting that simulated emitted power does not display significant seasonal variations either. The comparison between the simulated emitted power and the best measurement based on Voyager observations (Pearl et al., 1990) shows that the difference between simulation and measurement is smaller than the uncertainty of the measurement, suggesting that the simulation and measurement are statistically consistent.

**S6. REB During the Period of 1946-2030**

The absorbed solar power and emitted thermal power discussed in the previous sections are used to determine the REB at both global and hemispheric scales during the 1946-2030 period. The seasonal variations of the global-average radiant energy cycle (Fig. 3 in the main text) and its application to determine the internal heat are discussed in the main text. Here, we focus on the hemispheric-average REB, which will be used to determine the energy budget of Uranus' weather layer by combining internal heat with radiant energy components at the hemispheric scale (see Fig. 4 in the main text).

The unique obliquity of Uranus' rotation axis (97.7°) results in much stronger seasonal variations in hemispheric-average solar flux compared to those in global-average solar flux. Figure

S4 in the SI suggests that the seasonal variations of hemispheric-average solar flux can approach 100%, significantly larger than the seasonal variations of global-average solar flux (~17%). On the other hand, the albedo of solar-illuminated areas in each hemisphere (Fig. S11) exhibits much weaker seasonal variations, around 6%, which is extremely smaller than the seasonal variations in hemispheric-average solar flux (~100%). Therefore, the seasonal variations of hemispheric-average absorbed solar power, which closely follow the temporal patterns of hemispheric-average solar flux, are extremely strong. Panels A and B of Fig. S18 present the extremely strong seasonal variations of hemispheric-average absorbed solar power for the two hemispheres, respectively. On the other hand, the hemispheric-averaged emitted power remains statistically the same between the two hemispheres (Fig. S17) and constant over time. Panels C and D of Fig. S18 show the difference between hemispheric-average absorbed and emitted powers, indicating that the difference varies from a maximal energy excess (i.e., absorbed power > emitted power) with a value of 0.58±0.03 Wm$^{-2}$ at the summer solstice to a maximal energy deficit (i.e., absorbed power < emitted power) with a value of -0.68±0.02 Wm$^{-2}$ at the winter solstice of each hemisphere. Panels E and F of Fig. S18 show that the ratio of the hemispheric-average power difference to the emitted power changes from 84.0±4.6% at the summer solstice to -98.8±2.9% at the winter solstice for each hemisphere.

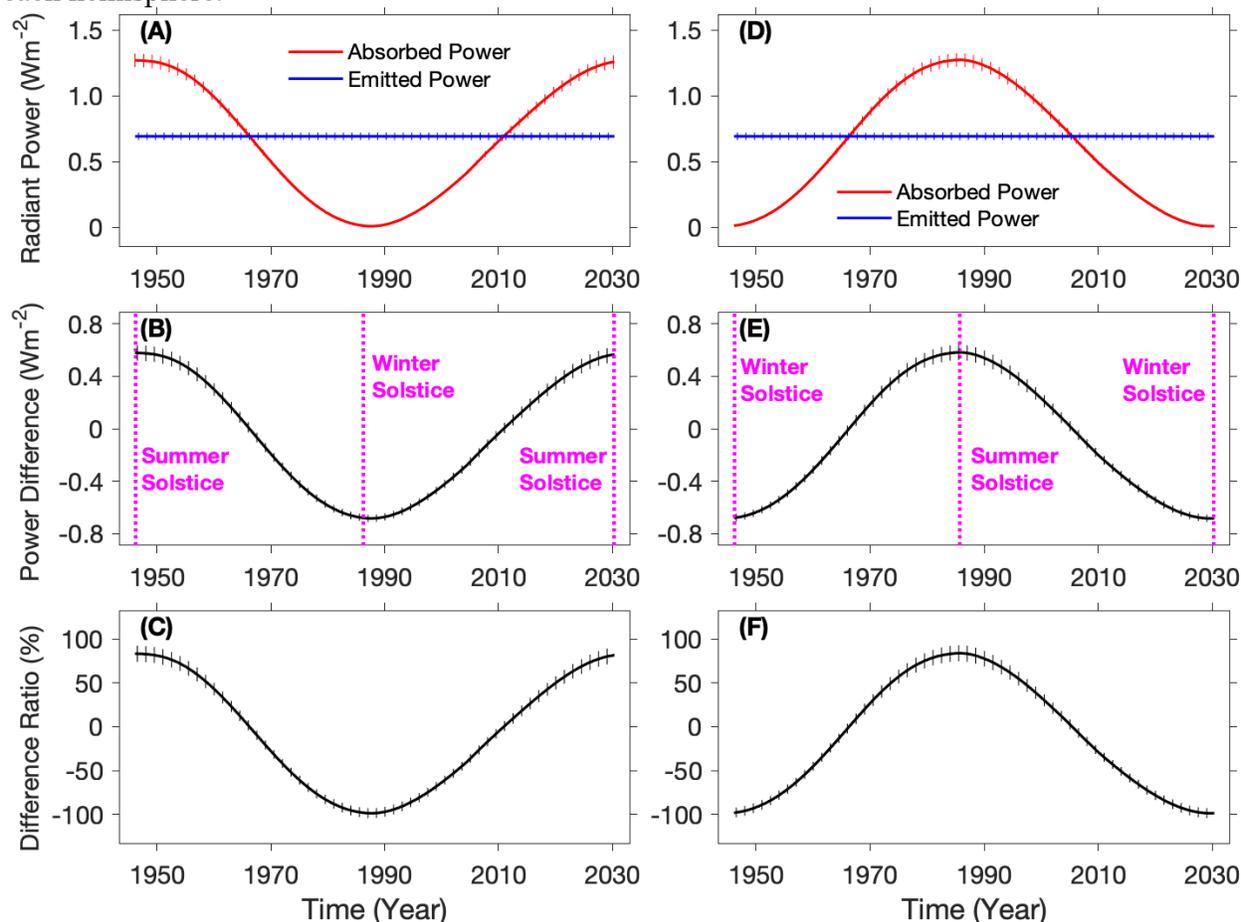

***Figure S18. Uranus' hemispheric-average REB and the difference between the two energy components (i.e., absorbed solar power and emitted thermal power).*** *(A) Comparison between the NH-average absorbed and emitted powers. (B) Difference between the NH-average absorbed and emitted powers (i.e., absorbed power minus emitted power). (C) Ratio of the difference to the*

*emitted power for the results shown in panel (B). Panels D, E, and F are the same as panels A, B, and C, respectively, except for SH-average analyses. The vertical solid thin lines in all six panels represent uncertainties. The three magenta vertical dashed lines in panels (B) and (E) represent the summer solstices of each hemisphere.*

**SI References**
**(References cited in the main text are not repeated here)**

Allison, M. and Travis, L. D., 1986. Astronomical, physical, and meteorological parameters for planetary atmospheres. The Jovian Atmospheres. NASA Conference Publication 2441.

Bevington, P. R. & Robinson, D. K., 2003. Data Reduction and Error Analysis for the Physical Sciences, 3rd ed., McGraw-Hill.

Caldwell, J., Wagener, R. and Fricke, K. H., 1988. Observations of Neptune and Uranus below 2000 Å with the IUE. Icarus 74, 133-140.

Cochran, W. D., Wagener, R., Caldwell, J. and Fricke, K. H., 1990. The ultraviolet continuum albedo of Uranus. Icarus 83, 93-101.

Courtin, R., Gautier, D. and Lacombe, A., 1978. On the thermal structure of Uranus from infrared measurements. Astronomy and Astrophysics 63, 97-101.

Creecy, E., Li, L., Jiang, X., Smith, M. D., Kleinboehl, A., Kass, D. M., and Martinez, G., 2022. Mars' Emitted Energy and Seasonal Energy Imbalance. PNAS, doi: 10.1073/pnas.2121084119.

Encrenaz, T., Schulz, B., Drossart, P., Lellouch, E., Feuchtgruber, H. and Atreya, S. K., 2000. The ISO spectra of Uranus and Neptune between 2.5 and 4.2 μm: constraints on albedos and H3+. Astronomy and Astrophysics 358, L83-L87.

Fazio, G.G., Traub, W.A., Wright, E.L., Low, F.J. and Trafton, L., 1976. Effective Temperature of Uranus. The Astrophysical Journal 209, 633-637.

Flasar, F. M. et al., 2004. Exploring the Saturn system in the thermal infrared: The Composite Infrared Spectrometer. Space Sci. Rev. 115, 169.

Hanel, R., Crosby, D., Herath, L., Vanous, D., Collins, D., Creswick, H., Harris, C. and Rhodes, M., 1980. Infrared spectrometer for Voyager. Applied Optics 19, 1391-1400.

Hapke, B., 2002. Bidirectional reflectance spectroscopy: 5. The coherent backscatter opposition effect and anisotropic scattering. Icarus 157, 523-534.

Henyey, L. G. and Greenstein, J. L., 1941. Diffuse radiation in the galaxy. Astrophysical Journal 93, 70-83.

Hildebrand, R. H., Loewenstein, R. F., Harper, D. A., Orton, G. S., Keene, J. and Whitcomb, S. E., 1985. Far-infrared and submillimeter brightness temperatures of the giant planets. Icarus 64, 64-87.

Karkoschka, E., 1998. Methane, Ammonia, and Temperature Measurements of the Jovian Planets and Titan from CCD–Spectrophotometry. Icarus 133, 134-146.

Li, L., B. Conrath, P. Gierasch, R. Achterberg, C. A. Nixon, A. A. Simon-Miller, F. M. Flasar, D. Banfield, K. H. Baines, R. A. West, A. R. Vasavada, A. Mamoutkine, M. Segura, G. Bjoraker, G. S. Orton, L. N. Fletcher, P. Irwin, P. Read, 2010. Emitted power of Saturn. J. Geophys. Res. 115, E11002, doi:10.1029/2010JE003631.

Li, L., K. H. Baines, M. A. Smith, R. A. West, S. Pérez-Hoyos, H. J. Trammell, A. Simon-Miller, B. Conrath, P. J. Gierasch, G. S. Orton, C. A. Nixon, G. Filacchione, P. M. Fry, and T. W. Momary, 2012. Emitted power of Jupiter based on Cassini CIRS and VIMS observations. J. Geophys. Res., doi:10.1029/2012JE004191.

Li, L., L. Guan, K. Heng, P. M. Fry, E. C. Creecy, R. J. Albright, T. D. Karandana, R. A. West, C. A. Nixon, M. E. Kenyon, A. Hendrix, U. Dyudina, 2023. The bolometric Bond albedo of Enceladus, Icarus 394, 115429.

Lockwood, G.W., Lutz, B.L., Thompson, D.T. and Warnock, A., 1983. The albedo of Uranus. Astrophysical Journal 266, 402-414.

Loewenstein, R. F., Harper, D. A., Moseley, S. H., Telesco, C. M., Thronson Jr, H. A., Hildebrand, R. H., Whitcomb, S. E., Winston, R. and Stiening, R. F., 1977. Far-infrared and submillimeter observations of the planets. Icarus 31, 315-324.

Moseley, H., Conrath, B. and Silverberg, R. F., 1985. Atmospheric temperature profiles of Uranus and Neptune. Astrophysical Journal 292, L83-L86.

Neff, J. S., Ellis, T. A., Apt, J. and Bergstralh, J. T., 1985. Bolometric albedos of Titan, Uranus, and Neptune. Icarus 62, 425-432.

Orton, G. S., 1985. A Pre-Voyager Infrared Perspective of the Atmosphere of Uranus. In Bulletin of the American Astronomical Society 17, 745.

Stier, M. T., Traub, W. A., Fazio, G. G., Wright, E. L. and Low, F. L., 1978. Far-infrared observations of Uranus, Neptune, and Ceres. Astrophysical Journal 226, 347-349.

Smith, B. A., Briggs, G. A., Danielson, G. E., Cook, A. F., Davies, M. E., Hunt, G. E., Masursky, H., Soderblom, L. A., Owen, T. C., Sagan, C. and Suomi, V. E., 1977. Voyager imaging experiment. Space Science Reviews 21, 103-127.

Wagener, R., Caldwell, J. and Fricke, K. H., 1986. The geometric albedos of Uranus and Neptune between 2100 and 3350 Å. Icarus 67, 281-288.

Wenkert, D., Deen, R. G. and Bunch, W. L., 2022, December. Uranus and Neptune Albedos May Need Revision. AGU Fall Meeting Abstracts, P32E-1865.

Yelle, R. V., McConnell, J. C., Strobel, D. F. and Doose, L. R., 1989. The far ultraviolet reflection spectrum of Uranus: Results from the Voyager encounter. Icarus 77, 439-456.